\documentclass[11pt]{article}

\usepackage{amsmath,amssymb}
\usepackage{graphicx}
\usepackage{amscd}
\usepackage{theorem}

\newcommand{\R}{{\mathbb{R}}}
\newcommand{\Z}{{\mathbb{Z}}}
\newcommand{\C}{{\mathbb{C}}}
\newcommand{\T}{{\mathbb{T}}}

\newcommand{\ba}{\begin{array}}
\newcommand{\ea}{\end{array}}

\newcommand{\bp}{\begin{pmatrix}}
\newcommand{\ep}{\end{pmatrix}}

\newcommand{\bps}{\begin{smallmatrix}}
\newcommand{\eps}{\end{smallmatrix}}

\newcommand{\ti}{\tilde}

\def \({\left(}
\def \){\right)}

\def \cA{{\cal A}}
\def \cB{{\cal B}}
\def \cC{{\cal C}}
\def \cD{{\cal D}}
\def \cE{{\cal E}}

\def \cL{{\cal L}}

\def \cS{{\cal S}}

\def \cV{{\cal V}}

\DeclareMathOperator{\Tr}{Tr}

\def \qed{\hfill $\blacksquare$}

\def \raw{\rightarrow}

\def \Im{\mathrm{Im}}

\def \dim{\mathrm{dim}}

\def \Hom{\mathrm{Hom}}
\def \End{\mathrm{End}}
\def \Ob{\mathrm{Ob}}

\def \Ker{\mathrm{Ker}}

\def \cCpre{\cC^{pre}_{\theta,E}}
\def \cCp0{\cC^{pre}_{\theta=0,E}}

\def \Hompre{\mathrm{Hom}_{\cC_{\theta,E}^{pre}}}
\def \Homp0{\mathrm{Hom}_{\cC_{\theta=0,E}^{pre}}}
\def \HomC{\mathrm{Hom}_{\cC}}

\def \ov#1{\frac{1}{#1}}

\def \zb{\bar{z}}

\def \delm#1{{\delta_{[#1]}}}

\def \bpart{\bar{\partial}}

\def \lpartial#1{\overleftarrow{\partial_{#1}}}
\def \rpartial#1{\overrightarrow{\partial_{#1}}}

\def \fpartial#1{\frac{\partial}{\partial {#1}}}

\def \g{{\frak g}}
\def \l{{\frak l}}

\def \S{{\frak S}}

\def \l({\left(}
\def \r){\right)}

\def \0{{\bf 0}}
\def \1{{\bf 1}} 

\def \Mat{\mathit{Mat}}
\def \ie{{\it i.e.\ }}

\def \fh{{\hat f}}

\def \Th{{\hat \T}}

\def \nabb{{\bar \nabla}}

\def \ib{{\bar i}}

\def \2i{\frac{\sqrt{-1}}{2}}
\def \i{\sqrt{-1}}

{
 \newtheorem{thm}{Theorem}[section]
 \newtheorem{prop}[thm]{Proposition}
 \newtheorem{lem}[thm]{Lemma}

\theorembodyfont{\rmfamily}
 \newtheorem{defn}[thm]{Definition}
 \newtheorem{rem}[thm]{Remark}
}

\begin{document}

\begin{titlepage}
\thispagestyle{empty}
\begin{flushleft}
\hfill YITP-05-63 \\
\hfill October, 2005
\end{flushleft}

\vskip 2.5 cm

\begin{center}

\noindent{\Large \textbf{Categories of 
holomorphic line bundles on higher dimensional noncommutative 
complex tori}}\\

\vspace*{0.8cm}

\noindent{
 }\\
\renewcommand{\thefootnote}{\fnsymbol{footnote}}

{\large 
Hiroshige Kajiura 
\footnote{e-mail address: kajiura@yukawa.kyoto-u.ac.jp}\\

\noindent{ \bigskip }\\

\it
Yukawa Institute for Theoretical Physics, Kyoto University \\
Kyoto 606-8502, Japan\\
}

\bigskip
\end{center}
\begin{abstract}

We construct explicitly noncommutative deformations of 
categories of holomorphic line bundles over higher dimensional tori. 
Our basic tools are Heisenberg modules over noncommutative tori 
and complex/holomorphic structures on them introduced by A.~Schwarz. 
We obtain differential graded (DG) categories 
as full subcategories of curved DG categories of Heisenberg modules 
over the complex noncommutative tori. 
Also, we present the explicit composition formula of morphisms, 
which in fact depends on the noncommutativity.

\end{abstract}

\vfill

\end{titlepage}
\vfill
\setcounter{footnote}{0}
\renewcommand{\thefootnote}{\arabic{footnote}}
\newpage

\tableofcontents

\section{Introduction}

In this paper, we propose a way to construct 
differential graded (DG) categories of 
finitely generated projective modules 
over higher dimensional noncommutative complex tori. 
Also, we give explicit examples of this construction 
as noncommutative deformations of the DG categories 
of holomorphic line bundles over higher dimensional 
complex tori. 

The motivation of the present paper is 
to construct an explicit example of the noncommutative 
deformations of a complex manifold which may be thought of as 
one of the extended deformations of a complex manifold 
proposed by Barannikov-Kontsevich \cite{BK}. 
For a $n$-dimensional complex or a Calabi-Yau manifold $M$, 
the extended deformation is defined 
by a deformation $\epsilon\in\g^1$ 
of the Dolbeault operator 
$\bpart:\g^k\to\g^{k+1}$ such that 
$(\bpart+\epsilon)^2=0$, where $\g:=\oplus_{k=0}^n\g^k$ 
is the graded vector space given by 
$\g^k:=\oplus_{k=p+q}
\Gamma(M,\wedge^p TM\otimes
\wedge^q {\bar TM}^*)$. 
The degree one graded piece consists of 
$\g^1=\Gamma(M,\wedge^2 TM)\oplus
\Gamma(M,TM\otimes {\bar TM}^*)\oplus
\Gamma(M,\wedge^2 {\bar TM}^*)$. 
Namely, it defines an extended deformation of the usual 
complex structure deformation 
$\epsilon\in\Gamma(M,TM\otimes {\bar TM}^*)$. 
In particular, the deformation corresponding to 
$\epsilon\in\Gamma(M,\wedge^2 TM)$ 
is called the {\em noncommutative deformation} 
of the complex manifold $M$. 
On the other hand, 
examples of the models of noncommutative deformation 
should be constructed so that 
we can see how it is noncommutative, 
as in the case of the deformation quantization \cite{Ko}. 
A candidate of it may be to consider an algebra deformation of $M$. 
Let $V:=\oplus_{k=0}^n V^k$ be a graded vector space given by 
$V^k:=\Gamma(M,\wedge^k {\bar TM})$. 
This $V$ has a natural graded commutative product 
$\cdot: V^k\otimes V^l\to V^{k+l}$ together with a differential 
given by the Dolbeault operator 
$\bpart:V^k\to V^{k+1}$. Then, 
$(V,\bpart,\cdot)$ forms a DG algebra. 
A deformation of this DG algebraic structure may describe 
the noncommutative deformation corresponding to 
$\epsilon\in\Gamma(M,\wedge^2 TM)$. 
We can also replace this DG algebra 
with the DG category of holomorphic vector bundles or 
coherent sheaves on $M$, where 
the DG algebra $V$ should be included in the DG category 
as the endomorphism algebra of the structure sheaf on $M$. 
This algebraic or categorical approach 
can be thought of as a part of the spirits of 
{\em homological mirror symmetry} by Kontsevich \cite{mirror}. 

Actually, (topological) string theory suggests 
considering the graded vector spaces $\g$ and $V$; 
the algebraic structures on their cohomologies are defined as 
the {\em closed {\em \cite{W1}} and open {\em \cite{W2}} string amplitudes}, 
respectively, in the B-twisted topological string theory (B-model). 
The DG category above corresponds to 
physically what is called a {\em D-brane category} (see \cite{Dbcat}); 
the objects are the D-branes, the morphisms are the open strings 
between the D-branes in the B-model. 
Thus, the approach above is physically 
to construct an open string model instead of the closed string model. 

It is natural from viewpoints 
of both string theory (see \cite{infty,thesis}) and deformation theory 
(see \cite{SS,FOOO} and references therein) that 
DG categories constructed as above should be treated 
in the category of $A_\infty$-categories, 
where equivalence between $A_\infty$-categories 
should be defined by the homotopy equivalence. 

Now, let us concentrate on a $n$-dimensional complex tori 
$(M:=T^{2n},\bpart)$. 
It would be easy if its noncommutative deformation corresponding to 
$\epsilon\in\Gamma(T^{2n},\wedge^2 TM)$ can be described by 
the DG algebra $(V,\bpart,*)$, where 
$ * :V^k\otimes V^l\to V^{k+l}$ is the natural extension 
of the Moyal product on $V^0$, the space of functions on $T^{2n}$, 
defined by the Poisson bivector $\epsilon\in\Gamma(T^{2n},\wedge^2 TM)$. 
However, as far as one identifies homotopy equivalent DG algebras 
with each other, 
all these DG algebras turn out to be equivalent, 
being independent of the noncommutative parameter $\epsilon$. 
In fact, one can easily show that the DG algebra $(V,\bpart,*)$ is 
formal, \ie, homotopy equivalent to 
a graded algebra on the cohomology $H(V,\bpart)$, 
and in particular the product on the cohomology $H(V,\bpart)$ 
is independent of $\epsilon$. 
These results follow from the fact that 
one can take a Hodge-Kodaira decomposition of the complex $(V,\bpart)$ 
so that the harmonic form is closed with respect to the product $*$. 

Therefore, in the same way as in the complex one-tori 
(= real two-tori) case \cite{foliation,PoSc,KimKim,nchms}, 
we should include nontrivial vector bundles 
which are compatible with the complex structure 
in some sense. 
In the real two-tori case, 
one can construct a DG category of 
holomorphic vector bundles, in the sense of \cite{Stheta}, 
over a noncommutative two-torus \cite{PoSc, nchms}, 
where holomorphic vector bundles are described 
by DG-modules. 
In particular, the derived category of the DG-category is 
independent of the noncommutativity parameter 
$\theta\in\R$ \cite{PoSc}. 
Though 
noncommutative deformation of complex tori in this approach is 
relatively well understood for complex one-tori case, 
its higher dimensional extension is quite nontrivial and interesting 
especially 
from the viewpoint of the extended deformation \cite{BK,Kap,gual,oren}. 
However, in this higher dimensional case, a different problem will
arise. Even though we start with a DG-module of 
a holomorphic vector bundle, 
its noncommutative deformation might not be described by a 
DG module. There may be several ways to resolve this problem. 
Our idea in this paper is that we treat the deformed holomorphic
vector bundles as {\em curved differential graded (CDG)}-modules 
over $V$. 
The important point is that even though the deformed ones are not 
DG modules, the space of morphisms may be equipped with a 
differential. 
Namely, in the context of DG categories, 
the cohomology should be defined not on the objects but 
on the {\em morphisms} between the objects. 
Thus, one may be able to 
extract finite dimensional graded vector spaces as 
the cohomologies of the morphisms. 

According to such a spirit, 
we construct DG categories 
consisting of some of these CDG modules 
of deformed holomorphic vector bundles 
on higher dimensional noncommutative tori. 
We remark that this procedure is just the same as the 
DG categories of B-twisted topological Landau-Ginzburg model 
by \cite{kl:1,kl:2,at} and also 
similar to the procedure by FOOO \cite{FOOO} in the mirror dual 
A-model side. 
It would also be interesting to construct a triangulated category 
via the twisted complexes as is done in \cite{at, at-wdg, Block}. 

Our starting point is based on A.~Schwarz's framework of 
noncommutative supergeometry \cite{S-Qalg} and 
noncommutative complex tori \cite{Stheta,Stensor}. 
In \cite{S-Qalg}, a CDG-algebra \cite{Pos} is re-studied and applied 
to noncommutative geometry under the name a {\em Q-algebra}, 
where modules over a Q-algebra is discussed. 
On the other hand, in \cite{Stheta}, a complex structure is introduced 
on a real $2n$-dimensional noncommutative torus $T^{2n}_\theta$, 
and a holomorphic structure on 
the {\em Heisenberg modules}, noncommutative analogs of vector
bundles, over $T^{2n}_\theta$ is defined. 
Then, our set-up can be thought of as 
an application of the noncommutative supergeometry 
\cite{S-Qalg} 
to the theory of holomorphic Heisenberg modules \cite{Stheta}. 
This set-up provides us with explicit descriptions of 
noncommutative models. 
Though one of our motivation comes from Fukaya's noncommutative model 
of Lagrangian foliations on symplectic tori \cite{f-nc} 
and their mirror dual, 
our approach in this paper is different from the one since 
we deal with the Heisenberg modules which are 
{\em finitely generated} projective modules 
over noncommutative tori. 
For recent papers, see \cite{BBK} for another approach 
to noncommutative complex tori and 
the set-up in \cite{Block} which should be closer to ours.

The construction of this paper is as follows. 
In section \ref{sec:CDG}, 
we recall the definitions of CDG algebras \cite{Pos}, CDG modules and 
CDG categories. 
The notion of modules over a Q-algebra is more general than that 
of the CDG-modules over a CDG algebra. 
However, for our purpose, it is enough to consider CDG modules 
since we begin with Heisenberg modules with a {\em constant} curvature 
connections. 
In section \ref{sec:CDGtori}, 
we construct CDG categories of Heisenberg modules over 
noncommutative tori with complex structures. 
In particular, we propose a way to obtain DG categories 
as full subcategories of the CDG categories. 
Along the general strategy in section \ref{sec:CDGtori}, 
we construct CDG categories of holomorphic line bundles over 
noncommutative complex tori and the DG categories 
as their full subcategories in section \ref{sec:three}. 
In subsection \ref{ssec:comm}, we construct the CDG category 
on a commutative complex tori. 
In this case, the CDG category is exactly a DG category. 
In subsection \ref{ssec:nc}, 
we consider three types of noncommutative deformations of the DG
category as CDG categories. 
Then, we obtain DG categories as the full subcategories of the 
CDG categories. 
Furthermore, we present the composition formula 
of the zero-th cohomologies of the DG categories explicitly. 
The structure constants of the compositions in fact 
depend on the noncommutative parameters, which implies that 
the DG categories or the triangulated/derived categories of them 
depend on the noncommutative parameters. 
These results 
can be thought of as generalizations of complex one-dimensional case 
\cite{foliation,PoSc,KimKim,nchms} and also 
a complex two-tori case \cite{KimLee,KimKim2} 
(in the case that the structure constant of the composition 
is not deformed by the noncommutative parameter). 
Also, from a string theory or homotopy algebraic point of view, 
these deformations should corresponds to 
deformations of an $A_\infty$-structure as weak $A_\infty$-algebras 
discussed in the context of open-closed homotopy algebras (OCHAs) 
\cite{ocha} (see also \cite{HLL}). 

{\bf Notations}: 
In this paper, any (graded) vector space stands for the one 
over the field $k=\C$. 
We use indices $i,j,\cdots$ for both the ones which run over
$1,\cdots,d=2n$ 
and the ones which run over $1,\cdots,n$, 
where $n$ and $d=2n$ are the complex and the real dimension 
of a noncommutative torus.

{\bf Acknowledgments}:\ 
I would like to thank O.~Ben-Bassat, A.~Kato, A.~Takahashi and 
Y.~Terashima 
for valuable discussions and useful comments. 
The author is supported by JSPS Research Fellowships for Young Scientists.

 \section{CDG algebras, CDG modules and CDG categories}
\label{sec:CDG}

\begin{defn}[(Cyclic) curved differential graded (CDG) algebra \cite{Pos}]
A {\em curved differential graded (CDG) algebra} 
$(V, f, d, m)$ consists of 
a $\Z$ (or $\Z_2$) graded vector space $V=\oplus_{k\in\Z} V^k$, 
where $V^k$ is the degree $k$ graded piece, 
equipped with a degree two element $f\in V^2$, a degree one differential 
$d :V^{k}\raw V^{k+1}$ and a degree preserving bilinear map 
$m : V^k\otimes V^{l}\raw V^{k+l}$ 
satisfying the following relations: 
\begin{align}
 &  d(f)=0\ ,\label{df} \\
 & (d)^2(v)= m(f,v) - m(v,f)\ ,\label{d2=f}\\
 &  d m(v,v')=m(d(v),v') 
 +(-1)^{|v|} m(v,d(v'))\ , \label{leib}\\
 & m(m(v,v'),v'') = m(v,m(v',v''))\ , \label{asso}
\end{align}
where $|v|$ is the degree of $v$, that is, $v\in V^{|v|}$. 

Suppose that we have in addition a nondegenerate symmetric inner product 
$$\eta:V^k\otimes V^l\raw \C$$ 
of fixed degree $|\eta|\in\Z$ on $V$. Namely, the $\eta$ 
is nondegenerate, nonzero only if $k+l+|\eta|=0$, and satisfies 
$\eta(v,v')=(-1)^{kl}\eta(v',v)$ for $v\in V^k$ and $v'\in V^l$. 
Then, we call $(V,f,\eta,d,m)$ 
a {\it cyclic CDG algebra} if the following conditions hold: 
\begin{align*}
 &\eta(d(v),v')+(-1)^{|v|}\eta(v,d(v'))=0\ ,
 &\eta(m(v,v'),v'')
 =(-1)^{|v|(|v'|+|v''|)}
 \eta(m(v',v''),v)\ .
\end{align*}
 \label{defn:CDG}
\end{defn}
\begin{rem}
A CDG algebra is identified with a weak $A_\infty$-algebra 
$(V,\{m_k:V^{\otimes k}\to V\}_{k\ge 0})$ 
with $m_0=f$, $m_1=d$, $m_2=\cdot$ and $m_3=m_4=\cdots=0$. 
This algebraic structure is what is called a Q-algebra 
introduced in the framework of 
noncommutative supergeometry in \cite{S-Qalg}. 
Also, a CDG algebra $(V,f,d,\cdot)$ with $f=0$ is a DG algebra, 
which is a (strict) $A_\infty$-algebra 
$(V,\{m_k\}_{k\ge 1})$ with $m_3=m_4=\cdots =0$. 
 \label{rem:CDG}
\end{rem}
\begin{defn}[CDG module]
A {\em right CDG module} $(\cE,d^\cE,m^\cE)$ over a CDG algebra $(V,-f,d,m)$ 
is a $\Z$-graded vector space $\cE$ 
equipped with a degree one linear map $d^\cE:\cE\to \cE$ and a right 
action $m^\cE:\cE\otimes V\to \cE$ 
satisfying the following condition: for any $v,v'\in V$ and 
$v^\cE\in\cE$, 
\begin{equation*}
 \begin{split}
  & (d^\cE)^2(v^\cE)=m^\cE(v^\cE, f)\ ,\\
  &  d^\cE m^\cE(v^\cE,v)=m^\cE(d^\cE(v^\cE),v)
 +(-1)^{|\xi|}m^\cE(v^\cE,d(v))\ ,\\
  & m^\cE(v^\cE,m(v,v'))=m^\cE(m^\cE(v^\cE,v),v')\ .
 \end{split}
\end{equation*}
In particular, if $f=0$, then $(\cE,d^\cE,m^\cE)$ is called 
a {\em DG-module} over a DG algebra $(V,d,m)$. 
\end{defn}
The third condition is nothing but the condition that 
$\cE$ is a (graded) right module over $V$. 

In the category extension of CDG algebras, 
elements of a CDG algebra turns out to be morphisms in the 
CDG-category. 
\begin{defn}[(Cyclic) CDG category]
A {\em CDG category} $\cC$ consists of a set of objects 
$\Ob(\cC)=\{a, b, \cdots\}$, 
a $\Z$-graded vector space 
$V_{ab}=\oplus_{k\in\Z} V^k_{ab}$ for each two objects 
$a$, $b$ and the grading $k\in\Z$, 
$f_a: \C\raw V^2_{aa}$ for each $a$, 
a differential 
$d :V^k_{ab}\raw V^{k+1}_{ab}$ and 
a composition of morphisms 
$m: V^k_{bc}\otimes V^l_{ab}\to 
V^{k+l}_{ac}$ satisfying the following relations: 
\begin{align}
 &  d(f_a)=0\ ,\label{df-cat} \\
 & (d)^2(v_{ab})=m(f_b,v_{ab}) - m(v_{ab},f_a)\ ,\label{d2=f-cat}\\
 &  d m(v_{bc},v_{ab})=m(d(v_{bc}),v_{ab})
 +(-1)^{|v_{bc}|} m(v_{bc},d(v_{ab}))\ , \label{leib-cat}\\
 & m(m(v_{cd},v_{bc}),v_{ab}) = m(v_{cd},m(v_{bc},v_{ab}))\ , 
 \label{asso-cat}
\end{align}
where $|v_{ab}|$ is the $\Z$-grading of $v_{ab}$, that is, 
$v_{ab}\in V_{ab}^{|v_{ab}|}$. 

Let $\eta$ be a nondegenerate symmetric inner product 
of fixed degree $|\eta|\in\Z$ on $V:=\oplus_{a,b}V_{ab}$. 
Namely, for each $a$ and $b$, 
\begin{equation}
 \eta: V^k_{ba}\otimes V^l_{ab}\raw \C
\end{equation}
is nondegenerate, nonzero only if $k+l+|\eta|=0$, and satisfies 
$\eta(V^k_{ba},V^l_{ab})=(-1)^{kl}\eta(V^l_{ab},V^k_{ba})$. 
In this situation, we call a CDG category with inner product $\eta$ 
a {\it cyclic CDG category} $\cC$ if the following conditions hold: 
\begin{align}
 &\eta(dv_{ab},v_{ab})+(-1)^{|v_{ab}|}\eta(v_{ab},dv_{ab})=0\ ,
 \label{d-int}\\
 &\eta(m(v_{bc}\otimes v_{ab}),v_{ca})
 =(-1)^{{|v_{bc}|}(|v_{ab}|+|v_{ca}|)}
 \eta(m(v_{ab}\otimes v_{ca}),v_{bc})\ .
 \label{cyclic}
\end{align}
Also, we call a cyclic CDG category 
$\cC$ a {\it cyclic DG category} 
if $f_a=0$ for any $a\in\Ob(\cC)$. 
 \label{defn:cDGcat}
\end{defn}
A CDG category $\cC$ consisting of one object only 
is a CDG algebra. 
Similarly, for a fixed object $a\in\Ob(\cC)$, 
the CDG category structure of $\cC$ reduces to 
a CDG algebra $(V_{aa},f_a,d,m)$. 
On the other hand, if the space of morphisms $\cV=\oplus_{a,b}V_{ab}$ 
is thought of as a $\Z$-graded vector space, 
$(\cV,\eta,\oplus_{a\in\Ob(\cC)}f_a,d,m)$ can be regarded as 
a cyclic CDG algebra (see also \cite{S-Qalg}). 

Suppose that a CDG category $\cC$ has an object 
$o\in\Ob(\cC)$ such that $(V_{oo},\eta,d,m)$ forms a DG algebra 
and for any object $a\in\Ob(\cC)$ 
there exists a center ${\hat f_a}$ in $V_{oo}$ such that 
\begin{equation*}
 m(f_a,v_{oa})=m(v_{oa},{\hat f_a})\ .
\end{equation*}
Then, $(V_{oa}=:\cE_a,d,m)$ can be regarded as a CDG module over the 
cyclic CDG algebra $(V_{oo},\eta,-{\hat f_a},d,m)$.

 \section{CDG modules and CDG categories on 
noncommutative tori}
\label{sec:CDGtori}

 \subsection{Higher dimensional noncommutative tori}
\label{ssec:nctori}

Let us consider an algebra generated by 
$U_i$, $i=1,\cdots, d$, with relations 
\begin{equation} \label{ui}
 U_jU_k = e^{-2\pi\i\theta^{jk}}U_{k}U_{j} \ ,\qquad  j,k=1,\cdots, d\ 
\end{equation}
for an antisymmetric $d\times d$ matrix 
$\theta:=\{\theta^{jk}\}$. 
Any element is spanned over $\C$ by elements 
$U_{\vec{m}}$, ${\vec{m}}= (m_1,\dots, m_d)\in\Z^d$, 
which are defined by 
\begin{equation*}
 U_{\vec{m}}:= U_{1}^{m_{1}}U_{2}^{m_{2}}
\dots U_{d}^{m_{d}}e^{\pi\i\sum_{j<k} m_{j}m_{k}\theta^{jk}}\ . 
\end{equation*}
The relation between $U_{\vec{m}}$ and $U_{\vec{m}'}$ becomes 
\begin{equation} \label{un}
U_{\vec{m}}U_{\vec{m}'} 
= e^{\pi\i\sum_{j,k}m_{j}\theta^{jk}m'_{k}} U_{\vec{m}+\vec{m}'}\ .
\end{equation}
One can represent any element of this algebra as 
\begin{equation*}
 u=\sum_{\vec{m}\in\Z^d} 
 u_{\vec{m}}U_{\vec{m}}\ ,\qquad u_{\vec{m}}\in\C\ .
\end{equation*}
For any element $u$ represented as above, 
an involution $*$ is defined by 
\begin{equation*}
 u^*:=\sum_{\vec{m}\in\Z^d} u_{\vec{m}}^* U_{\vec{m}}^*\ ,
\end{equation*}
where $u_{\vec{m}}^*$ is the complex conjugate of $u_{\vec{m}}$ and 
$U_{\vec{m}}^{*}:= U_{-\vec{m}}$. 
One can consider a subalgebra $T_\theta^d$ such that any 
element, again represented as $u=\sum_{{\vec{m}}} u_{\vec{m}}U_{\vec{m}}$, 
belongs to the Schwartz space $\cS(\Z^d)$, that is, 
the coefficients $\{u_{\vec{m}}\}$ as a function on $\Z^d$ tend to 
zero faster than any power of $||{\vec{m}}||$. 
This algebra $T_\theta^d$ is in fact a $C^*$-algebra and called 
(the smooth version of) a {\em noncommutative torus} 
\cite{Rhigh,KS}. 

There is a canonical normalized {\em trace} on $T_{\theta}^{d}$  
specified by the rule 
\begin{equation} \label{Tr}
 \Tr (u)=u_{\vec{m}=0}\ ,\qquad u=\sum_{{\vec{m}}} u_{\vec{m}}U_{\vec{m}}\ .
\end{equation}
For $\theta = 0$ we can realize the algebra $T_{\theta}^{d}$ 
as an algebra of 
functions on a $d$-dimensional torus $T^{d}$. 
Then the trace (\ref{Tr}) corresponds 
to an integral over $T^{d}$ provided the volume of $T^{d}$ is one.

In order to define a connection on a module $E$ over 
noncommutative torus $T_{\theta}^{d}$ we 
shall first define a natural Lie algebra of shifts $\cL_\theta$ 
acting on $T_{\theta}^{d}$. 
The shortest way to define this Lie algebra is 
by specifying a basis consisting of derivations 
$\delta_j:T^d_\theta\to T^d_\theta$, $j=1,\dots, d$, 
satisfying 
\begin{equation} \label{delta}
\delta_{j} (U_{\vec{m}}) = 2\pi\i m_{j}U_{\vec{m}} \, . 
\end{equation}
For the multiplicative generators $U_{j}$ the above relation reads as 
\begin{equation*} \label{deltaij}
\delta_{j}U_{k} = 2\pi\i \delta_{jk}U_{k} \, .
\end{equation*}
This derivations then span a $d$-dimensional abelian Lie algebra 
(over $\C$) 
that we denote $\cL_\theta$. 

A {\em connection} on a (right) module $E$ over $T_{\theta}^d$ is 
a set of operators $\nabla_{X}:E\to E$, $X\in\cL_\theta$ 
depending linearly on $X$ and satisfying 
\begin{equation*}
 \nabla_X(\xi\cdot u)=\nabla_X(\xi)\cdot u+\xi\cdot X(u)\ 
\end{equation*}
for any $\xi\in E$ and $u\in T_{\theta}^{d}$. 
In general, connections whose curvature equals 
the identity endomorphism times a numerical tensor are called 
constant curvature connections. 
We express the {\em constant curvature} of 
a constant curvature
connection $\nabla_i:=\nabla_{\delta_i}$, $i=1,\cdots, d$, as 
\begin{equation}\label{F}
 [\nabla_i,\nabla_j]
=-2\pi\sqrt{-1}F_{ij}\cdot\1_{\End_{T_{\theta}^{d}}(E)}\ , \qquad 
 F_{ij}=-F_{ji}\in\R\ .
\end{equation}

On a noncommutative torus, one can construct a class of 
finitely generated projective modules called 
{\em Heisenberg modules} (see \cite{Rhigh, KS}). 
In fact, 
on any Heisenberg module 
there exists a constant curvature connection. 
They play a special role. 
It was shown by Rieffel (\cite{Rhigh}) 
that if the matrix $\theta^{ij}$ is irrational in the sense 
that at least one of its entries is irrational then 
any projective module over $T_{\theta}^{d}$ 
is isomorphic to a direct sum of Heisenberg modules. 

Heisenberg modules are applied to discuss the Morita equivalence of 
noncommutative tori. 
A noncommutative tori $T^d_\theta$ is Morita equivalent to
$T^d_{\theta'}$ if \cite{RS} and only if \cite{S} 
$\theta'=g(\theta)$, $g\in SO(d,d,\Z)$ 
(for more recent papers, see \cite{TW,hLi,Ell-Li}). 
Here, the $SO(d,d,\Z)$ action on the space of skew symmetric matrices 
in $\Mat_d(\R)$ is defined by 
\begin{equation*}
 g(\theta):= (\cA\theta+\cB)(\cC\theta+\cD)^{-1}\ ,
\quad 
 g:=\bp \cA & \cB \\ \cC & \cD \ep \in SO(d,d,\Z)\ ,
\end{equation*}
where 
$SO(d,d,\Z):=\{g\in\Mat_{2n}(\Z)\ |\ g^tJg=J \}$ 
for 
$J:=\left(\bps \0_n & \1_n \\ \1_n & \0_n\eps\right)$. 
To establish this Morita equivalence, one may construct a 
$T^d_{\theta}$-$T^d_{g(\theta)}$ {\em Morita equivalence bimodule}, 
denoted by $P_{\theta\text{-}g(\theta)}$ (see \cite{Rhigh,RS}). 
One can in fact construct 
the Morita equivalence bimodule $P_{\theta\text{-}g(\theta)}$ 
for any $g\in SO(d,d,\Z)$ 
as a left Heisenberg module $E$ over $T^d_\theta$. 
In this case, the algebra $\End_{T^d_\theta}(E)$, 
the algebra of endomorphisms of $E$ which commute with the 
left action of $T_\theta^d$, 
coincides with the noncommutative torus $T^d_{g(\theta)}$. 
This implies that one can construct a 
$T^d_{g(\theta)}$-$T^d_{\theta}$ Morita equivalence bimodule 
$P_{g(\theta)\text{-}\theta}$ 
as a right Heisenberg module over $T^d_\theta$. 
We denote it by $E_{g,\theta}$; 
we have $\End_{T^d_\theta}(E_{g,\theta})\simeq T_{g(\theta)}^d$. 
Also, the $T^d_\theta$-$T^d_{g(\theta)}$ Morita equivalence bimodule
is given by the right Heisenberg module $E_{g^{-1},g(\theta)}$. 
On $T_{g(\theta)}^d$, 
a {\em trace} $\Tr_{T_{g(\theta)}^d}:T_{g(\theta)}^d\to\C$ and 
{\em derivations} $\delta_i:T_{g(\theta)}^d\to T_{g(\theta)}^d$, 
$i=1,\cdots, d$, are defined by appropriate rescaling of those for 
$T^d_\theta$ as 
\begin{equation}\label{Tr-End}
 \Tr_{T_{g(\theta)}^d}(u)
 =\sqrt{|\det(\cC\theta+\cD)|}\, u_{{\vec{m}}=0}
 \,\qquad 
 u:=\sum_{{\vec{m}}\in\Z^n}
 u_{\vec{m}} Z_{{\vec{m}}}\in T_{g(\theta)}^d\ 
\end{equation}
and 
\begin{equation*}
 \delta_j(Z_{{\vec{m}}})
 =\frac{2\pi\i m_j}{\sqrt{|\det(\cC\theta+\cD)|}}\, Z_{{\vec{m}}}\ ,
\end{equation*}
where $g=\left(\bps\cA & \cB \\ \cC & \cD\eps\right)$ and 
$Z_1,\cdots,Z_d$ are the generators of $T^d_{g(\theta)}$ 
with relations 
$Z_j Z_k = e^{-2\pi\i(g\theta)^{jk}}Z_k Z_j$, $j,k=1,\cdots, d$. 
For $X\in\cL_\theta$, a linear map 
$\nabla:T_{g(\theta)}^d\otimes\cL_\theta\to T_{g(\theta)}^d$ is 
defined by extending linearly 
\begin{equation}\label{delta-End}
 \nabla_{\delta_i}(u):=\delta_i(u)\ ,\quad 
 i=1,\cdots,d\ ,\qquad u\in T_{g(\theta)}^d\ .
\end{equation}

When we have a $T^d_\theta$-$T^d_{\theta'}$ Morita equivalence
bimodule, one can consider the following {\em tensor product} 
(see \cite{S,S-Qalg}): 
\begin{equation}
 E_{g_a,\theta}\otimes_{T^d_\theta}P_{\theta\text{-}g(\theta)}
 \simeq  E_{g_a g^{-1},g(\theta)}\ 
 \label{MEtensor}
\end{equation}
for a Heisenberg module $E_{g_a,\theta}$ with any $g_a\in  SO(d,d,\Z)$, 
where the tensor product $\otimes_{T^d_\theta}$ is defined by 
the standard tensor product $\otimes$ over $\C$ with the identification 
$(\xi_a\cdot u)\otimes p\sim \xi_a\otimes (u\cdot p)$ 
for any $\xi_a\in E_{g_a,\theta}$, $u\in T_\theta^d$ and 
$p\in P_{\theta\text{-}g(\theta)}$. 
Let us denote $\theta_a:=g_a(\theta)$. 
For a right Heisenberg module 
$E_{g_a,\theta}$ and a Morita equivalence bimodule 
$P_{\theta_b\text{-}\theta_a}$ with 
a $SO(d,d,\Z)$ element $g_b$, the existence of the tensor product 
(\ref{MEtensor}) implies that 
we have the following tensor product: 
\begin{equation*}
P_{\theta_b\text{-}\theta_a}\otimes_{T_{\theta_a}^d}E_{g_a,\theta}
 \simeq E_{g_b,\theta}\ .
\end{equation*}
Thus, we may identify $P_{\theta_b\text{-}\theta_a}$ with 
the space of homomorphisms from $E_{g_a,\theta}$ to $E_{g_b,\theta}$. 
Hereafter we write 
$$\Hom(E_{g_b,\theta},E_{g_a,\theta})
:=P_{g_b(\theta)\text{-}g_a(\theta)}\ .$$ 

On a Morita equivalence $T^d_{\theta_b}$-$T^d_{\theta_a}$ bimodule 
$\Hom(E_{g_a,\theta},E_{g_b,\theta})$, 
we define a connection 
$\nabla:\Hom(E_{g_a,\theta},E_{g_b,\theta})\otimes\cL_\theta
\to \Hom(E_{g_a,\theta},E_{g_b,\theta})$ 
by a linear map satisfying the following relation: 
\begin{equation*}
  \nabla_X(u_b\cdot \xi)=\nabla_X(u_b)\cdot\xi
  +u_b\cdot\nabla_X(\xi)\ ,\quad  
  \nabla_X(\xi\cdot u_a)= \nabla_X(\xi)\cdot u_a+\xi\cdot\nabla_X(u_a)\ ,
 \quad 
 u_a\in T_{\theta_a}^d\ ,\quad
 u_b\in T_{\theta_b}^d\ ,
\end{equation*}
where $\nabla_X(u_b)$ and $\nabla_X(u_a)$ 
are defined by eq.(\ref{delta-End}). 
Since these Morita equivalence bimodules are Heisenberg modules, 
they can be equipped with constant curvature connections. 

A Heisenberg module $E_{g,\theta}$ attached to an element 
$g\in SO(d,d,\Z)$ as above is called a {\em basic module}. 
Since any Heisenberg module is constructed by a direct sum of 
basic modules, in this paper we concentrate on categories of 
basic modules. 
Let $\Ob:=\{a,b,\cdots\}$ be a collection of labels 
and consider a map $g:\Ob\to SO(d,d,\Z)$, $g(a):=g_a\in SO(d,d,\Z)$ for 
$a\in\Ob$. 
For the collection of Heisenberg modules 
$\{E_{g_a,\theta}\ |\ a\in\Ob\}$, 
assume we have an associative {\rm product} 
\begin{equation*}
 m: \Hom(E_{g_b,\theta},E_{g_c,\theta})\otimes
\Hom(E_{g_a,\theta},E_{g_b,\theta})
\to\Hom(E_{g_a,\theta},E_{g_c,\theta})
\end{equation*}
for any $a,b,c\in\Ob$. 
Namely, for any $a,b,c,d\in\Ob$ and 
$\xi_{ab}\in\Hom(E_{g_a,\theta},E_{g_b,\theta})$, 
$\xi_{bc}\in\Hom(E_{g_b,\theta},E_{g_c,\theta})$, 
$\xi_{cd}\in\Hom(E_{g_c,\theta},E_{g_d,\theta})$, 
we assume that the product $m$ satisfies 
\begin{equation}\label{nc-associativity}
 m(m(\xi_{cd},\xi_{bc}),\xi_{ab})
 =m(\xi_{cd},m(\xi_{bc},\xi_{ab}))\ .
\end{equation}
Such a product 
$m: \Hom(E_{g_b,\theta},E_{g_c,\theta})\otimes
\Hom(E_{g_a,\theta},E_{g_b,\theta})
\to\Hom(E_{g_a,\theta},E_{g_c,\theta})$ is essentially 
the tensor product; $m$ is 
constructed by fixing a map inducing the isomorphism 
\begin{equation*}
 \Hom(E_{g_b,\theta},E_{g_c,\theta})\otimes_{T^d_{\theta_b}}
\Hom(E_{g_a,\theta},E_{g_b,\theta})
 \simeq\Hom(E_{g_a,\theta},E_{g_c,\theta})\ .
\end{equation*}
There exists a choice of the map so that the associativity
holds (see \cite{S-Qalg}). 

For $a\in\Ob$, suppose that 
a constant curvature connection $\nabla_a$ is defined on
$E_{g_a,\theta}$. 
Also, for $b\in\Ob$, $a\ne b$, 
define a constant curvature connection 
$\nabla:\Hom(E_{g_a,\theta},E_{g_b,\theta})\otimes\cL_\theta\to
\Hom(E_{g_a,\theta},E_{g_b,\theta})$ whose 
constant curvature $F_{ab}:=\{F_{ab,ij}\}_{i,j=1,\cdots,d}$ is defined by 
\begin{equation*}
 F_{ab,ij}\cdot\xi_{ab}:=\frac{\i}{2\pi}[\nabla_i,\nabla_j](\xi_{ab})
 \ ,\qquad F_{ab,ij}=-F_{ab,ji}\in\R
\end{equation*}
for any $\xi\in\Hom(E_{g_a,\theta},E_{g_b,\theta})$. 
Then, a constant curvature connection 
$\nabla_b:E_{g_b,\theta}\otimes\cL_\theta\to E_{g_b,\theta}$ 
can be induced as follows \cite{S,S-Qalg}:
\begin{equation*}
 \nabla_b(m(\xi_{ab},\xi_a)):=
 m(\nabla(\xi_{ab}),\xi_a)+m(\xi_{ab},\nabla_a(\xi_a))\ 
\end{equation*} 
for any $\xi_a\in E_{g_a,\theta}$ and 
$\xi_{ab}\in\Hom(E_{g_a,\theta},E_{g_b,\theta})$, 
where the relation between the curvatures of $E_{g_a,\theta}$, 
$E_{g_b,\theta}$ and that of $\Hom(E_{g_a,\theta},E_{g_b,\theta})$ 
is given by 
\begin{equation*}
 F_b-F_a=F_{ab}\ .
\end{equation*}
Thus, repeating this procedure leads to the following category 
$\cC_{\theta,E}^{pre}$: 
\begin{defn}\label{defn:pre-C}
For a noncommutative torus $T^d_\theta$, let 
$\Ob:=\{a,b,\cdots\}$ be a collection of labels and 
$E$ a map from $\Ob$ to the space of basic modules with constant
curvature connections; for $a\in\Ob$, we denote 
$E(a)=(E_{g_a,\theta},\nabla_a)=:E_a$. 
A {\em category} $\cC_{\theta,E}^{pre}$ is defined by the following data.

$\bullet$\ The collection of objects is 
\begin{equation*}
 \Ob(\cC_{\theta,E}^{pre}):=\Ob\ .
\end{equation*}
Each object $a\in\Ob$ is associated with a basic module with a 
constant curvature connection $E_a$ whose constant curvature is
denoted by a skewsymmetric matrix $F_a\in\Mat_{2n}(\R)$.

$\bullet$\ 
For any $a,b\in\Ob(\cC_{\theta,E}^{pre})$, 
the space of morphisms is 
\begin{equation*}
 \Hom_{\cC_{\theta,E}^{pre}}(a,b):=\Hom(E_{g_a,\theta},E_{g_b,\theta})\ ,
\end{equation*}
which is equipped with a constant curvature connection 
$\nabla:\Hompre(a,b)\otimes\cL_\theta\to
\Hompre(a,b)$ with its constant curvature $F_{ab}=F_b-F_a$. 

$\bullet$\ 
For any $a,b,c\in\Ob$, there exists an associative product 
(eq.(\ref{nc-associativity}))
\begin{equation*}
 m:\Hom_{\cC_{\theta,E}^{pre}}(b,c)\otimes
\Hom_{\cC_{\theta,E}^{pre}}(a,b)\to\Hom_{\cC_{\theta,E}^{pre}}(a,c)\ .
\end{equation*}

$\bullet$\ 
For any $a,b,c\in\Ob$ and 
$\xi_{ab}\in\Hom_{\cC_{\theta,E}^{pre}}(a,b)$, 
$\xi_{bc}\in\Hom_{\cC_{\theta,E}^{pre}}(b,c)$, 
the Leibniz rule holds: 
\begin{equation}\label{nc-leibniz}
 \nabla m(\xi_{bc},\xi_{ab}) = 
  m(\nabla(\xi_{bc}),\xi_{ab}) 
  +m(\xi_{bc},\nabla(\xi_{ab}))\ .
\end{equation}

$\bullet$\ 
For any $a\in\Ob$, 
a trace $\Tr_a:\Hompre(a,a)\to\C$ 
is given by eq.(\ref{Tr-End}): 
\begin{equation*}
 \Tr_a (u)
 =\sqrt{|\det(\cC_a\theta+\cD_a)|}\, u_{{\vec{m}}=0}
 \,\qquad 
 u:=\sum_{{\vec{m}}\in\Z^n}
 u_{\vec{m}} Z_{{\vec{m}}}\in\Hompre(a,a)\ 
\end{equation*}
for $g_a=\left(\bps\cA_a & \cB_a \\ \cC_a & \cD_a\eps\right)$. 
In particular, for any $a,b\in\Ob$, 
$\Tr_a m: 
\Hom_{\cC_{\theta,E}^{pre}}(b,a)\otimes
\Hom_{\cC_{\theta,E}^{pre}}(a,b)\to\C$ gives a nondegenerate 
bilinear map such that 
\begin{equation}\label{pre-cyclic}
 \Tr_a m(\xi_{ba},\xi_{ab})=\Tr_b m(\xi_{ab},\xi_{ba})\ ,\qquad 
 \xi_{ab}\in\Hom_{\cC_{\theta,E}^{pre}}(a,b)\ ,\quad 
 \xi_{ba}\in\Hom_{\cC_{\theta,E}^{pre}}(b,a)\ .
\end{equation}
\end{defn}
The last identity (\ref{pre-cyclic}) 
together with the nondegeneracy of $\Tr_a m$ 
is a typical property of 
Morita equivalence bimodules (see \cite{Rhigh,KS}, 
and for noncommutative two-tori case \cite{nchms}).

 \subsection{Noncommutative complex tori and CDG structures on them}
\label{ssec:nctoriCDG}

Let us consider a complex structure on the noncommutative torus 
$T_\theta^{2n}$ as introduced by A.~Schwarz \cite{Stheta}. 
We take a different notation which fits our arguments, 
though it is equivalent to the one in \cite{Stheta}. 
When we define a complex structure on a commutative torus $T^{2n}$, 
we may take a $\C$-valued $n$ by $n$ matrix 
$\tau=\{\tau_j^{\, i}\}$, 
$i,j=1,\cdots,n$, whose imaginary part $\tau_I:=\Im(\tau)$ 
is positive definite. 
A commutative complex torus is then described by
$\C^n/(\Z^n+\tau^t\Z^n)$, 
where $\tau^t$ is the transpose of $\tau$. 
The complex coordinates of $\C^n$ are given by $(z_1,\cdots,z_n)$, 
$z^i=x^i+\sum_j y^j\tau_j^{\, i}$, $i=1,\cdots, n$. 
The corresponding Dolbeault operator $\bpart$ is given by 
\begin{equation*}
 \bpart=\sum_{i=1}^nd\zb^i\fpartial{\zb^i}\ ,\quad 
  \fpartial{\zb^i}
 :=\ov{2\sqrt{-1}}\sum_{j=1}^n
 \(((\tau_I)^{-1}\tau)_i^j\fpartial{x^j}-((\tau_I)^{-1})_i^j\fpartial{y^j}\)\ ,
\end{equation*}
where we denote $\Im(\tau)=:\tau_I$ which is 
by definition positive definite.

Based on these formula, for 
a noncommutative torus $T^{2n}_\theta$ and 
a fixed complex structure $\tau$, 
let us define $\bpart_i\in\cL_\theta$, $i=1,\cdots, n$, by 
\begin{equation*}
 \bpart_i:=\ov{2\sqrt{-1}}\sum_{j=1}^n
 \(((\tau_I)^{-1}\tau)_i^j\delta_j-((\tau_I)^{-1})_i^j\delta_{n+j}\)\ .
\end{equation*}
Also, for $E_a:=(E_{g_a,\theta},\nabla_a)$, 
a Heisenberg module $E_{g_a,\theta}$ over $T^{2n}_\theta$ with a 
constant curvature connection $\nabla_{a,i}$, $i=1,\cdots, 2n$, 
define a {\em holomorphic structure} 
$\nabb_{a,i}:E_{g_a,\theta}\to E_{g_a,\theta}$, $i=1,\cdots,n$, by 
\begin{equation}\label{hol}
 \nabb_{a,i}:=\ov{2\sqrt{-1}}\sum_{j=1}^n
 \(((\tau_I)^{-1}\tau)_i^j\nabla_{a,j}-((\tau_I)^{-1})_i^j\nabla_{a,n+j}\)\ .
\end{equation}
For each pair $(E_a,E_b)$, we define 
a {\em holomorphic structure} 
$\nabb_i:\Hom(E_{g_a,\theta},E_{g_b,\theta})
\to\Hom(E_{g_a,\theta},E_{g_b,\theta})$, $i=1,\cdots,n$, 
by the same formula: 
\begin{equation}\label{hol-hom}
 \nabb_{i}:=\ov{2\sqrt{-1}}\sum_{j=1}^n
 \(((\tau_I)^{-1}\tau)_i^j\nabla_{j}-((\tau_I)^{-1})_i^j\nabla_{n+j}\)\ .
\end{equation}
Let $\Lambda$ be the Grassmann algebra generated by 
$d\zb^1,\cdots,d\zb^n$ of degree one. 
Namely, they satisfy $d\zb^id\zb^j=-d\zb^jd\zb^i$ for any
$i,j=1,\cdots,n$, so in particular $(d\zb^i)^2=0$. 
These generators are thought of as the formal basis 
of the antiholomorphic one forms on the complex noncommutative torus 
$T_\theta^{2n}$. 
By $\Lambda^k$ we denote the degree $k$ graded piece of $\Lambda$. 
The graded vector space $V:=T^{2n}_\theta\otimes\Lambda$ 
is then thought of as the space of the smooth 
antiholomorphic forms on the complex noncommutative torus 
$T_\theta^{2n}$, which also has the 
the graded piece decomposition: 
\begin{equation*}
  V=\oplus_{k=0}^n V^k\ .
\end{equation*}
Any element in $V^k$ can be written as 
\begin{equation*}
 v=\sum_{{\vec{m}}\in\Z^n}\sum_{i_1,\cdots,i_k}
 v_{{\vec{m}};i_1\cdots i_k}U_{{\vec{m}}}\cdot 
   (d\zb^{i_1}\cdots d\zb^{i_k})\ ,
\end{equation*}
where $v_{{\vec{m}};i_1\cdots i_k}\in\C$ is 
skewsymmetric with respect to the indices $i_1\cdots i_k$. 
A product $m:V^k\otimes V^l\to V^{k+l}$ 
is defined naturally by combining the product on $T_{\theta}^{2n}$ 
with the one on the Grassmann algebra $\Lambda$, and then 
$(V,m)$ forms a graded algebra. 
One can define a differential $d:V^k\to V^{k+1}$, 
\begin{equation*}
 d:=\sum_{i=1}^nd\zb^i\cdot\bpart_i\ , 
\end{equation*}
which satisfies the Leibniz rule with respect to 
the product $m$. 

An inner product $\eta: V^k\otimes V^l\raw\C$ 
of degree $-n$ is defined by 
the composition of the product $m$ with a trace 
$\int_{T_\theta^{2n}} :V\raw\C$: 
\begin{equation*}
 \eta=\int_{T_\theta^{2n}} m\ ,\qquad 
 \int_{T_\theta^{2n}} v = v_{{\vec{m}}=0;i_1\cdots i_k}
 \epsilon^{i_1\cdots i_k}_{1\cdots n}\ . 
\end{equation*}
Here $\epsilon$ is defined by 
\begin{equation*}
 \epsilon^{i_1\cdots i_k}_{1\cdots n}=
 \begin{cases}
  0 &  (k\ne n) \\
 \sum_{\sigma\in\S_n}\epsilon(\sigma)\delta^{i_1}_{\sigma(1)}\cdots
 \delta^{i_n}_{\sigma(n)}& k=n \ ,
 \end{cases}
\end{equation*}
where $\epsilon(\sigma)$ is the signature of the permutation 
of $n$ elements $\sigma\in\S_n$. 
Namely, $\int_{T_\theta^d} : V^k\raw\C$ 
is thought of as the integration of differential forms 
over $T^{2n}_\theta$, 
as an extension of the trace map $\Tr:T_\theta^{2n}\to\C$ 
in eq.(\ref{Tr}), and hence gives a nonzero map only if $k=n$. 
\begin{lem}
$(V,\eta,d,m)$ forms a cyclic DG algebra. 
\qed\end{lem}
For $E_a:=(E_{g_a,\theta},\nabla_a)$ a Heisenberg module 
over $T_{\theta}^{2n}$ with a 
constant curvature connection, 
we lift $E_a$ to a $\Z$-graded right module 
$\cE_a:=E_{g_a,\theta}\otimes\Lambda$ over $V$. 
We denote by $m_a:\cE_a\otimes V\to\cE_a$ 
the right action of $V$ on $\cE_a$. 
The connection $\nabla_a:E_{g_a,\theta}\otimes\cL_\theta\to
E_{g_a,\theta}$ is then lifted to 
a degree one linear map $d_a:\cE_a\to\cE_a$ defined by 
\begin{equation*}
 d_a:=\sum_{i=1}^nd\zb^i\cdot\nabb_{a,i} \ ,
\end{equation*}
where $\nabb_{a,i}$ is the holomorphic structure (\ref{hol}). 
This $d_a$ is not a differential in general. 
Namely, the graded module has its curvature:
\begin{equation*}
 (d_a)^2 v^{\cE_a}= \fh_a\cdot v^{\cE_a}\ ,\qquad 
 \fh_a:=-\pi\i 
 \left( d\zb^t\tau_I^{-1}\right)
 \bp \tau & -\1_n\ep F_a 
 \bp \tau^t \\ -\1_n \ep 
 \left(\tau_I^{t,-1}d\zb \right)\in\Lambda^2\ 
\end{equation*}
for any $v^{\cE_a}\in\cE_a$, where $d\zb^t:=(\zb^1,\cdots,\zb^n)$. 
This $d_a$ defines a differential on $\cE_a$, 
that is, $\fh_a=0$ if and only if 
\begin{equation*}
 \bp \tau & -\1_n \ep F_a\bp \tau^t \\ -\1_n \ep=0\ .
\end{equation*}
In this case, $(\cE_a,d_a,m_a)$ forms a DG module over $V$. 
In the commutative case ($\theta=0$), 
this condition on $F_a$ is nothing but the condition that 
the corresponding (line) bundle is holomorphic, 
\ie, the curvature is of $(1,1)$-form with respect to the complex 
structure defined by $\tau$. 
However, for general $\theta$, $\fh_a$ can not be zero even if 
it is zero when $\theta$ is set to be zero. 

On the other hand, 
since $\fh_a\in\Lambda^2\subset V^2$ is a center in $V$ 
with respect to the product $m$, 
$(d)^2(v)=m(\fh_a,v)-m(v,\fh_a)\ (=0)$ holds and then 
$(V,\eta,-\fh_a,d,m)$ forms a cyclic CDG algebra. Thus: 
\begin{lem}\label{lem:H-CDG}
$(\cE_a, \fh_a, d_a,m_a)$ forms a CDG module over 
the cyclic CDG algebra $(V,\eta,-\fh_a,d,m)$. 
\qed\end{lem}
We call this $\fh_a\in V^2$ the {\it potential two-form} of $\cE_a$.

Now, for a category $\cC^{pre}_{\theta,E}$ given in 
Definition \ref{defn:pre-C}, 
we construct a CDG category $\cC_{\theta,\tau,E}=:\cC$. 
\begin{defn}\label{defn:C}
For a fixed category $\cC^{pre}_{\theta,E}$, 
a {\em category} $\cC$ is defined as follows. 

$\bullet$\ 
The collection of objects is 
\begin{equation*}
 \Ob(\cC):=\Ob\ ,
\end{equation*}
where any object $a\in\Ob$ is associated with a CDG module 
$(\cE_a,\fh_a,d_a,m_a)$ over the CDG algebra 
$(V,\eta,-\fh_a,d,m)$ corresponding to $E_a$ as in 
Lemma \ref{lem:H-CDG}. 

$\bullet$\ 
For any $a,b\in\Ob(\cC)$, the space of morphisms is 
the graded vector space 
\begin{equation*}
 \HomC(a,b):=\Hompre(a,b)\otimes\Lambda=:V_{ab}=\oplus_{k=1}^n V_{ab}^k\ ,
\end{equation*}
which is equipped with 
a degree one linear map $d:V_{ab}^k\to V_{ab}^{k+1}$, 
\begin{equation*}
 d:=\sum_{i=1}^nd\zb^i\nabb_i\ ,
\end{equation*}
where $\nabb_i$ is the holomorphic structure (\ref{hol-hom}) 
corresponding to the constant curvature connection 
$\nabla:\Hompre(a,b)\otimes\cL_\theta\to\Hompre(a,b)$. 

$\bullet$\ For any $a,b,c\in\Ob(\cC)$, an associative product 
$m:V_{bc}^l\otimes V_{ab}^k\to V_{ac}^{k+l}$ is given by 
the lift of the product $m$ on $\Hompre(*,*)$. 

$\bullet$\ 
For any two objects $a,b\in\Ob(\cC)$, 
a nondegenerate graded symmetric inner product 
$\eta: V_{ba}\otimes V_{ab}\to\C$ of degree $-n$ 
is defined by 
\begin{equation*}
 \eta=\int_{T_{\theta_a}^{2n}} m\ ,\qquad 
 m:V_{ba}\otimes V_{ab}\to V_{aa}\ .
\end{equation*}
Here $\int_{T_{\theta_a}^{2n}}:V_{aa}\to\C$ is defined by 
\begin{equation*}
 \int_{T_{\theta_a}^{2n}} v
 =\sqrt{|\det(\cC_a\theta+\cD_a)|}\, v_{{\vec{m}}=0;i_1\cdots i_k}
 \epsilon^{i_1\cdots i_k}_{1\cdots n}\  \,\quad 
  v:=\sum_{{\vec{m}}\in\Z^n}\sum_{i_1,\cdots,i_k}
 v_{{\vec{m}};i_1\cdots i_k}U_{{\vec{m}}}\cdot 
   (d\zb^{i_1}\cdots d\zb^{i_k})\in V_{aa}\ 
\end{equation*}
for $g_a=\left(\bps\cA_a & \cB_a \\ \cC_a & \cD_a\eps\right)$ 
as an extension of the trace map $\Tr_a\to\C$. 
\end{defn}
Due to the Leibniz rule (\ref{nc-leibniz}), 
it is clear that $d:V_{ab}^k\to V_{ab}^{k+1}$ is a 
derivation: 
\begin{equation}
 d m(v_{bc}\otimes v_{ab}) = 
 m(d(v_{bc})\otimes v_{ab}) 
 +(-1)^{|v_{bc}|}m(v_{bc}\otimes d(v_{ab}))\ .
 \label{v-Leibniz}
\end{equation}
Let us define $\fh_{ab}\in\Lambda^2$ by 
\begin{equation*}
 \fh_{ab}:=d^2\ ,\qquad d:V_{ab}^k\to V_{ab}^{k+1}\ .
\end{equation*}
Then, the Leibniz rule (\ref{v-Leibniz}) and $F_{ab}=F_b-F_a$ imply 
$\fh_{ab}=\fh_b-\fh_a$. 
For each $a\in\Ob(\cC)$, let $f_a:=\fh_a\cdot\1_a\in V_{aa}$, where 
$\1_a$ is the identity in $T^{2n}_{\theta_a}$. 
The following is the main claim of this paper. 
\begin{prop}[Cyclic DG category of holomorphic vector bundles]\hfill
\begin{itemize}
 \item[(i)]\ 
For a given category $\cC^{pre}_{\theta,E}$ in Definition
\ref{defn:pre-C}, $\cC:=\cC_{\theta,\tau,E}$ 
forms a cyclic CDG category. 
 \item[(ii)]\ 
Let $\cC^\fh$ be the full subcategory of $\cC$ such that 
any $a\in\Ob(\cC^\fh)\subset\Ob(\cC)$ 
is a set of CDG modules 
over a cyclic CDG algebra $(V,\eta,-\fh,d,m)$ for a fixed 
$\fh\in\Lambda^2\subset V^2$. 
Then $\cC^\fh$ forms a cyclic DG category. 
 \end{itemize}
 \label{prop:CDGcat}
\qed\end{prop}
This implies that one can construct a kind of DG categories of 
holomorphic vector bundles over a noncommutative tori.

 \section{Three examples of noncommutative deformations} 
\label{sec:three}

Now, we construct examples of various 
noncommutative deformations of the DG categories 
of Heisenberg modules described by CDG modules 
over a cyclic CDG algebra of a noncommutative torus. 
The set-up given in the previous subsection 
allows us to deform both the complex structure $\tau$ and 
the noncommutativity $\theta$ or either of them. 
In this paper, starting from a commutative ($\theta=0$) 
$n$ dimensional complex torus 
with the standard complex structure $\tau=\i\1_n$ in subsection 
\ref{ssec:comm}, 
we deform the noncommutative parameter $\theta$ with 
preserving the standard complex structure in subsection 
\ref{ssec:nc}. 
Also, we give the composition formula on the zero-th cohomologies 
of the DG categories explicitly. We show that 
the structure constants of the compositions in fact depends on 
$\theta$.

 \subsection{Commutative case}
\label{ssec:comm}

Let us begin with the commutative torus 
$T^{2n}:=T^{2n}_{\theta=0}$. 
The generators 
$U_1,\cdots,U_n,U_{{\bar 1}}:=U_{n+1},\cdots, U_{{\bar n}}:=U_{2n}$ 
then commute with each other. 
The arguments in subsection \ref{ssec:nctoriCDG} 
show that it is enough to construct a 
category $\cC^{pre}_{\theta=0,E}$ 
in order to construct a cyclic CDG category $\cC$. 

A category $\cC^{pre}_{\theta=0,E}$ is constructed as follows. 
Any object $a\in\Ob(\cC^{pre}_{\theta=0,E})$ 
is associated with a pair $E_a:=(E_{g_a,\theta=0},\nabla_a)$ of 
basic module $E_{g_a,\theta=0}$ with a 
constant curvature connection $\nabla_a$. 
The basic module is defined by 
$$E_{g_a,\theta=0}:=\cS(\R^n\times(\Z^n/A_{a}\Z^n))$$ 
for a fixed nondegenerate symmetric matrix $A_a\in\Mat_n(\Z)$, 
where $g_a\in SO(d,d,\Z)$ is given by 
\begin{equation*}
 g_a= \bp \1_{2n} & \0_{2n} \\ F_a & \1_{2n}\ep\ ,\qquad 
 F_a :=\bp \0_n & A_{a}\\ -A_{a} & \0_n\ep\ .
\end{equation*}
The right action of $T^{2n}$ on $E_{g_a,\theta=0}$ is defined by 
specifying the right action of each generator; 
for $\xi_a\in E_{g_a,\theta=0}$, it is given by 
\begin{equation*}
 \begin{split}
 (\xi_aU_i)(x;\mu)& =\xi_a(x;\mu) e^{2\pi\sqrt{-1}(x_i+(A_{a}^{-1}\mu)_i))} 
 \ ,\\
 (\xi_aU_\ib)(x;\mu)& =\xi_a(x+A_{a}^{-1}t_i;\mu-t_i)
 \ ,\qquad i=1,\cdots, n \ ,
 \end{split}
\end{equation*}
where $x:=(x_1\cdots x_n)^t\in\R^n$, $\mu\in\Z^n/A_{a}\Z^n$ 
and $t_i\in\R^n$ is defined by $(t_1 \cdots t_n)=\1_n$. 
A constant curvature connection 
$\nabla_a:E_{g_a,\theta=0}\otimes\cL_\theta\to E_{g_a,\theta=0}$ 
is given by 
\begin{equation}
 (\nabla_{a,1}\cdots \nabla_{a,2n})^t= 
 \bp
  \1_n &  \\ 
    & -A_{a} 
 \ep
 \bp \partial_x \\ 2\pi\sqrt{-1}x \ep\ ,
 \label{4dim-conn}
\end{equation}
where 
$\partial_x:=\left(\bps\fpartial{x_1}& \cdots &
\fpartial{x_n}\eps\right)^t$, 
and the curvature (defined by eq.(\ref{F})) is 
$F_a$ above. 
The generators of the endomorphism algebra is the same as $U_i,
U_\ib$:  
\begin{equation*}
 \begin{split}
 (Z_i\xi_a)(x;\mu)& 
 =e^{2\pi\sqrt{-1}(x_i+(A_{a}^{-1}\mu)_i))}\xi_a(x;\mu)\ ,\\
 (Z_\ib\xi_a)(x;\mu)& =\xi_a(x+A_{a}^{-1}t_i;\mu-t_i)
 \ ,\qquad i=1,\cdots, n \ .
 \end{split}
\end{equation*}
Namely, the endomorphism algebra also forms 
a commutative torus $T^{2n}$. 

This $E_a:=(E_{g_a},\nabla_a)$ is lifted 
to a CDG module $(\cE_a,\fh_a,d_a,m_a)$ over the 
cyclic CDG algebra $(V=T^{2n}\otimes\Lambda,\eta,-\fh_a,d,m)$ 
by the procedure in the previous subsection, where 
the complex structure is taken to be the standard one: 
$\tau=\i\1_n$. 
Then, one obtains $(d_a)^2=\fh_a=0$, that is, 
$\cE_a$ in fact forms a DG-module corresponding to 
a {\em holomorphic line} bundle. 

For any $a,b\in\Ob(\cC^{pre}_{\theta=0,E})$, 
the space $\Homp0(a,b)$ is defined as follows. 
If $A_{ab}:=A_b-A_a$ is nondegenerate, 
then it is again the Schwartz space
$\Homp0(a,b):=\cS(\R^n\times(\Z^n/A_{ab}\Z^n))$. 
For $\xi_{ab}\in\Homp0(a,b)$, the right action of $T^{2n}$, 
generated by $U_i$ and $U_\ib$, 
and the left action of $T^{2n}$, 
generated by $Z_i$ and $Z_\ib$, are defined by 
\begin{equation*}
  \begin{split}
 (\xi_{ab} U_i )(x;\mu)& = 
 \xi_{ab}(x;\mu) e^{2\pi\sqrt{-1}(x_i+(A_{ab}^{-1}\mu)_i))} \ ,\\
 (\xi_{ab} U_\ib)(x;\mu)& =\xi_{ab}(x+A_{ab}^{-1}t_i;\mu-t_i)\ ,\\
 (Z_i \xi_{ab})(x;\mu)& = e^{2\pi\sqrt{-1}(x_i+(A_{ab}^{-1}\mu)_i))} 
 \xi_{ab}(x;\mu)\ ,\\
 (Z_\ib \xi_{ab})(x;\mu)& =\xi_{ab}(x+A_{ab}^{-1}t_i;\mu-t_i)\ 
 \end{split}
\end{equation*}
for $i=1,\cdots,n$, where $\mu\in\Z^n/A_{ab}\Z^n$. 
In fact, all these generators $U_i$, $U_\ib$, $Z_i$ and $Z_\ib$ 
commute with each other. 

On the other hand, if $A_a=A_b$, 
we define $\Homp0(a,b):=T^{2n}$, 
on which the left and right actions of $T^{2n}$ are defined 
just by the commutative product on $T^{2n}$. 

In general, 
the way of constructing the space 
$\Hompre(a,b)=\Hom(E_{g_a,\theta=0},E_{g_b,\theta=0})$ 
depends on the rank of $A_{ab}:=A_b-A_a$. 
In rank $n$ case (nondegenerate case), 
one has $\Homp0(a,b):=\cS(\R^n\times(\Z^n/A_{ab}\Z^n))$ 
and in rank $0$ case, one has $\Homp0(a,b):=T^{2n}$ as above. 
In rank $1<r<n$ case, we should combine these two 
constructions with each other appropriately. 
In order to avoid such case-by-case arguments, 
in this paper we assume that 
$A_{ab}$ is nondegenerate for any
$a,b\in\Ob(\cC^{pre}_{\theta=0,E})$ such that $a\ne b$. 

The constant curvature connection 
$\nabla_i:\Homp0(a,b)\to\Homp0(a,b)$, 
$i=1,\cdots, 2n$, is given by 
\begin{equation*}
 (\nabla_1 \cdots \nabla_{2n})^t := 
 \bp
  \1_n &  \\ 
    & -A_{ab} 
 \ep
 \bp \partial_x \\ 2\pi\sqrt{-1}x \ep\ 
\end{equation*}
if $a\ne b$, and if $a=b$, it is defined by the 
derivation $\nabla$ 
on the noncommutative torus $T^{2n}_{\theta_a}=T^{2n}_{\theta_b}$ in 
eq.(\ref{delta-End}) with $\theta_a=\theta_b=0$. 

For $a,b,c\in\Ob(\cCp0)$ and 
$\xi_{ab}\in\Homp0(a,b)$, $\xi_{bc}\in\Homp0(b,c)$, 
the product 
$m:\Homp0(b,c)\otimes\Homp0(a,b)\to\Homp0(a,c)$ 
is given as follows:  

For $a=b$, it is the right action of $T^{2n}$ 
on $\Homp0(b,c)$. 

For $b=c$, it is the left action of $T^{2n}$ 
on $\Homp0(a,b)$. 

For $a=c$, the product 
$m:\Homp0(b,a)\otimes\Homp0(a,b)\to T^{2n}$ is given by 
\begin{equation*}
 m(\xi_{ba},\xi_{ab})(x,\rho)
 =\sum_{\vec{m}\in\Z^n}\sum_{\mu\in\Z^n/A_{ab}\Z^n} 
 U_{\vec{m}}\int_{\R^n} dx^n 
 \xi_{ba}(x,\mu)\cdot(\xi_{ab}(x,-\mu)U_{-\vec{m}})\ 
\end{equation*}
for $\xi_{ab}\in\Homp0(a,b)$ and $\xi_{ba}\in\Homp0(b,a)$, 
where $U_{\vec{m}}\in\Homp0(a,a)=T^{2n}$. 
For the remaining general case, it is given by 
\begin{equation}
 m(\xi_{bc},\xi_{ab})(x,\rho)
 =\sum_{u\in\Z^n}
  \xi_{bc}(x+A_{bc}^{-1}(u-A_{ab}A_{ac}^{-1}\rho),-u+\rho)\cdot
  \xi_{ab}(x-A_{ab}^{-1}(u-A_{ab}A_{ac}^{-1}\rho),u)\ .
 \label{tensor-comm}
\end{equation}
These structures together with the trace map 
defined by eq.(\ref{Tr}) forms a category $\cCp0$ 
and the corresponding cyclic CDG category $\cC_{\theta=0}$. 
In particular, we have $d^2=0$ for $d:V_{ab}\to V_{ab}$ 
with any pair $a,b\in\Ob(\cC_{\theta=0})$. 
Thus, $\cC_{\theta=0}$ is a cyclic DG category. 

For any $a,b\in\Ob(\cC_{\theta=0})$, $a\ne b$, 
the bases of the zero-th 
cohomology of $V_{ab}$ are given by gaussians 
\cite{Stheta} (see also \cite{DKL}) and called 
the {\em theta vector}, though here we are discussing the $\theta=0$ case. 
We shall give examples of these theta vectors in noncommutative case 
$\theta\ne 0$ in the next subsection. 
The mirror dual $\Th^{2n}$ of this complex torus 
$T^{2n}:=\C^n/(\Z^n\oplus \i\Z^n)$ 
is the real $2n$-dimensional torus with a symplectic structure 
$\omega:=\left(\bps \0_n & -\1_n \\ \1_n & \0_n \eps\right)$. 
In this mirror dual torus $\Th^{2n}$, a line bundle specified by 
$A_a$ corresponds to an affine lagrangian submanifold $L_a$. 
Then, the intersection of $L_a$ and $L_b$ is a point ${\hat v}_{ab}$ 
on $\Th^{2n}$, which defines 
the set $\ti{V}_{ab}$ of the infinite copies 
of the points on the covering space $\C^n$. 
The structure constant $C_{abc,\rho}^{\mu\nu}\in\C$ can be identified 
with the sum of the exponentials of the symplectic areas 
of the triangles 
$\ti{v}_{ab}\ti{v}_{bc}\ti{v}_{ac}$ for any 
$\ti{v}_{ab}\in\ti{V}_{ab}$, 
$\ti{v}_{bc}\in\ti{V}_{bc}$ and 
$\ti{v}_{ac}\in\ti{V}_{ac}$ with respect to $\omega$, 
where the triangles related by a parallel translations on 
the covering space are identified with each other and not
overcounted (see \cite{nctheta}).

 \subsection{Noncommutative deformations of holomorphic line bundles}
\label{ssec:nc}

Let us consider a real $2n$-dimensional noncommutative torus 
$T^{2n}_\theta$ with its generators 
$U_1,\cdots, U_{2n}$ with the following relation:
\begin{equation*}
 U_iU_j=e^{-2\pi\sqrt{-1}\theta^{ij}}U_jU_i\ ,\qquad 
 \theta:=\bp \theta_1 & -\theta_2 \\ \theta_2^t & \theta_3 \ep\ .
\end{equation*}
Since $\theta\in\Mat_{2n}(\R)$ is antisymmetric, 
$\theta_1,\theta_3\in\Mat_n(\R)$ are antisymmetric and 
$\theta_2\in\Mat_n(\R)$ can be an arbitrary $n$ by $n$ matrix.

A Heisenberg module $E_{g,\theta}$, $g\in SO(2n,2n,\Z)$, 
on this noncommutative torus $T^{2n}_\theta$ 
is associated with two notions, the $K_0$ group element and 
the Chern character (see \cite{KS}). 
The Chern character of $E_{g,\theta}$ is defined 
by its constant curvature 
$F$, a skewsymmetric $2n$ by $2n$ matrix with entries in $\R$. 
On the other hand, the $K_0$ group element of $E_{g,\theta}$ is defined 
by $F_0$, the constant curvature of $E_{g,\theta}$ when we set
$\theta=0$. 
Thus, the $K_0$ group element is independent of the 
noncommutativity $\theta$. 
A Heisenberg module $E_{g,\theta}$ 
is thought of as a noncommutative analog of a line bundle 
if $g\in SO(2n,2n,\Z)$ is of the form: 
\begin{equation*}
 g=\bp \1_{2n} & \0_{2n} \\ F_0 & \1_{2n}\ep\ 
\end{equation*}
for a skew symmetric matrix $F_0\in\Mat_{2n}(\Z)$. 
In fact, $F_0$ corresponds to the first Chern character of a 
line bundle if $\theta=0$. 
Since we shall discuss noncommutative deformations of the line bundles 
in the previous subsection, 
let us consider in particular the case that 
$F_0$ is of the following form
\begin{equation*}
 F_0=\bp \0_n & A \\ -A & \0_n \ep\ ,
\end{equation*}
where $A\in\Mat_n(\Z)$ is a nondegenerate symmetric matrix. 
For the Heisenberg modules $E_{g,\theta}$ with $g$ given as above, 
we shall consider noncommutative tori of the following three cases: 
\begin{center}
{\bf Type} $\theta_1$: $\theta_2=\theta_3=0$, \quad
{\bf Type} $\theta_2$: $\theta_1=\theta_3=0$, \quad 
{\bf Type} $\theta_3$: $\theta_1=\theta_2=0$\ . 
\end{center}
In each case, the endomorphism algebra, 
$T_{\theta'}$, 
$\theta':=(\1_n\theta + \0_n)(F_0\theta + \1_n)^{-1}$, 
turns out to be as follows. 
In Type $\theta_1$ case and $\theta_3$ case: 
\begin{equation*}
 \theta=\bp \theta_1 & \0_n \\ \0_n & \0_n\ep\ ,\qquad 
 \theta=\bp \0_n & \0_n \\ \0_n & \theta_3\ep\ ,
\end{equation*}
we have $\theta'=\theta$. 
However, in Type $\theta_2$ case, one obtains 
\begin{equation}
 \theta'=\bp \0_n & -\theta_2(\1_n+A\theta_2)^{-1} \\
             \theta_2^t(\1_n+A\theta_2^t)^{-1} & \0_n \ep\ ,
\qquad \theta:=\bp \0_n & -\theta_2 \\ \theta_2^t & \0_n\ep\ .
 \label{theta_2-transf}
\end{equation}

 \vspace*{0.3cm}

Now, for the cyclic DG category $\cC_{\theta=0}$ 
in the previous subsection, 
we construct its noncommutative deformations of three types above 
explicitly as CDG categories. 
Namely, we construct noncommutative deformations of 
$\cCp0$ in the previous subsection. 

\vspace*{0.2cm}

\noindent
{\bf Type $\theta_1$}\quad  
A category $\cCpre$ is constructed as follows. 
Any object $a\in\Ob(\cCpre)$ is associated with a pair 
$E_a:=(E_{g_a,\theta},\nabla_a)$. The basic module $E_{g_a,\theta}$ 
is defined by 
$$E_{g_a,\theta}=\cS(\R^n\times(\Z^n/A_a\Z^n))$$ 
for a nondegenerate symmetric matrix $A_a\in\Mat_n(\Z)$, where 
\begin{equation*}
 g_a= \bp \1_{2n} & \0_{2n} \\ F_{a,0} & \1_{2n}\ep\ ,\qquad 
 F_{a,0}:=\bp \0_n & A_{a}\\ -A_{a} & \0_n\ep\ .
\end{equation*}
For $\xi_a\in E_{g_a,\theta}$, 
the action of each generator is defined by 
\begin{equation*}
 \begin{split}
 (\xi_aU_i )(x;\mu)& = 
 \xi_a(x;\mu)* e^{2\pi\sqrt{-1}(x_i+(A_{a}^{-1}\mu)_i))} \ ,\\
 (\xi_aU_\ib)(x;\mu)& =\xi_a(x+A_{a}^{-1}t_i;\mu-t_i)
 \ ,\qquad i=1,\cdots, n \ . 
 \end{split}
\end{equation*}
Here $*: C^\infty(\R^n)\otimes C^\infty(\R^n)\to C^\infty(\R^n)$ 
is the Moyal star product (\cite{Moyal}) defined by 
\begin{equation*}
 (f * g)(x):=f(x)
e^{\frac{\sqrt{-1}}{4\pi}\lpartial{x}\theta_1\rpartial{x}}
 g(x)\ ,
\end{equation*}
where $\lpartial{x}\theta_1\rpartial{x}:=\sum_{p,q=1}^n
\lpartial{x_p}\theta^{pq}_1\rpartial{x_q}$. 
The generator of the endomorphism is then given by 
\begin{equation*}
 \begin{split}
 (Z_i\xi_a)(x;\mu)& = e^{2\pi\sqrt{-1}(x_i+(A_{a}^{-1}\mu)_i))} 
 * \xi_a(x;\mu)\ ,\\
 (Z_\ib\xi_a)(x;\mu)& =\xi_a(x+A_{a}^{-1}t_i;\mu-t_i)\ .
 \end{split}
\end{equation*}
A constant curvature connection 
$\nabla_a: E_{g_a,\theta}\otimes\cL_\theta\to E_{g_a,\theta}$ 
is given as 
\begin{equation*}
 (\nabla_{a,1},\cdots,\nabla_{a,n})^t = 
 \bp
  \1_n &  \\ 
  \ov{2}A_a\theta_1 & -A_a 
 \ep
 \bp \partial_x \\ 2\pi\sqrt{-1}x \ep\ ,
\end{equation*}
whose the constant curvature is 
\begin{equation*}
 F_a 
 = \bp \0_n & A_a\\ -A_a & -A_a\theta_1 A_a\ep\ .
\end{equation*}
We assume that 
$A_{ab}$ is nondegenerate for any $a,b\in\Ob(\cCpre)$, $a\ne b$. 
For any $a,b\in\Ob(\cCpre)$, 
the space $\Hompre(a,b)$ is defined as follows. 
If $a\ne b$, $\Hompre(a,b):=\Hom(E_{g_a,\theta},E_{g_b,\theta})
=\cS(\R^n\times(\Z^n/A_{ab}\Z^n))$; 
the right action of $T_{\theta_a}$, 
generated by $U_i$ and $U_\ib$, 
and the left action of $T_{\theta_b}$, 
generated by $Z_i$ and $Z_\ib$, are defined by 
\begin{equation*}
  \begin{split}
 (\xi_{ab} U_i )(x;\mu)& = 
 \xi_{ab}(x;\mu)* e^{2\pi\sqrt{-1}(x_i+(A_{ab}^{-1}\mu)_i))} \ ,\\
 (\xi_{ab} U_\ib)(x;\mu)& =\xi_{ab}(x+A_{ab}^{-1}t_i;\mu-t_i)\ ,\\
 (Z_i \xi_{ab})(x;\mu)& = e^{2\pi\sqrt{-1}(x_i+(A_{ab}^{-1}\mu)_i))} 
 * \xi_{ab}(x;\mu)\ ,\\
 (Z_\ib \xi_{ab})(x;\mu)& =\xi_{ab}(x+A_{ab}^{-1}t_i;\mu-t_i)\ .
 \end{split}
\end{equation*}
If $a=b$, then $\Hompre(a,b)=T_{\theta_a}=T_{\theta_b}$ and 
these actions are 
defined by the usual product of noncommutative torus 
$T_{\theta_a}=T_{\theta_b}$. 
The constant curvature connection 
$\nabla:\Hompre(a,b)\otimes\cL_\theta\to\Hompre(a,b)$ is given by 
\begin{equation*}
 (\nabla_{1}\cdots \nabla_{2n})^t:=\bp
  \1_n &  \\ 
  \ov{2}A_{ab}^+\theta_1 & -A_{ab} 
 \ep
 \bp \partial_x \\ 2\pi\sqrt{-1}x \ep \ ,
 \qquad A_{ab}^+:=A_a+A_b\ 
\end{equation*}
if $a\ne b$, and if $a=b$, it is defined by the 
derivation $\nabla$ of the noncommutative torus 
$T_{\theta_a}=T_{\theta_b}$ in eq.(\ref{delta-End}). 

For any $a,b,c\in\Ob(\cCpre)$ and 
$\xi_{ab}\in\Hompre(a,b)$, $\xi_{bc}\in\Hompre(b,c)$, 
the product 
$m:\Hompre(b,c)\otimes\Hompre(a,b)\to\Hompre(a,c)$ 
is given as follows:  

For $a=b$, it is the right action of $T_{\theta_a}=T_{\theta_b}$ 
on $\Hompre(b,c)$. 

For $b=c$, it is the left action of $T_{\theta_b}=T_{\theta_c}$ 
on $\Hompre(a,b)$. 

For $a=c$, the product 
$m:\Hompre(b,a)\otimes\Hompre(a,b)\to T_{\theta_a}$ is given by 
\begin{equation*}
 m(\xi_{ba},\xi_{ab})(x,\rho)
 =\sum_{\vec{m}\in\Z^n}\sum_{\mu\in\Z^n/A_{ab}\Z^n} 
 U_{\vec{m}}\int_{\R^n} dx^n 
 \xi_{ba}(x,\mu) * (\xi_{ab}(x,-\mu)U_{-\vec{m}})\ .
\end{equation*}
For the remaining general case, it is given by 
\begin{equation*}
 m(\xi_{bc},\xi_{ab})(x,\rho)
 =\sum_{u\in\Z^n}
  \xi_{bc}(x+A_{bc}^{-1}(u-A_{ab}A_{ac}^{-1}\rho),-u+\rho) * 
  \xi_{ab}(x-A_{ab}^{-1}(u-A_{ab}A_{ac}^{-1}\rho),u)\ .
\end{equation*}
The trace map is then given by eq.(\ref{Tr-End}). 

\vspace*{0.2cm}

\noindent
{\bf Type $\theta_2$}\quad  
A category $\cCpre$ is constructed as follows. 
Any object $a\in\Ob(\cCpre)$ is associated with a pair 
$E_a:=(E_{g_a,\theta},\nabla_a)$, 
where we assume $\det(\1_n+\theta_2 A_a)\ne 0$, 
which is always satisfied 
if one of the entries of $\theta_2$ is irrational. 
The basic module $E_{g_a,\theta}$ is defined by 
\begin{equation*}
E_{g_a,\theta}=\cS(\R^n\times(\Z^n/A_a\Z^n))
\end{equation*}
for a nondegenerate symmetric matrix $A_a\in\Mat_n(\Z)$, where 
\begin{equation*}
 g_a= \bp \1_{2n} & \0_{2n} \\ F_{a,0} & \1_{2n}\ep\ ,\qquad 
 F_{a,0} =\bp \0_n & A_{a}\\ -A_{a} & \0_n\ep\ .
\end{equation*}
For $\xi_a\in E_{g_a,\theta}$, 
the action of each generator is defined by 
\begin{equation*}
  \begin{split}
 (\xi_a U_i )(x;\mu)& = \xi_a(x;\mu)
 e^{2\pi\sqrt{-1}x_i+(A_a^{-1}\mu)_i))} \ ,\\
 (\xi_aU_\ib)(x;\mu)& 
 = \xi_a(x+(\1_n+\theta_2 A_a)A_a^{-1}t_i;\mu-t_i)\ .
 \end{split}
\end{equation*}
The action of the generators of the endomorphism is then given by 
\begin{equation*}
 \begin{split}
 (Z_i\xi_a)(x;\mu)& 
 = e^{2\pi\sqrt{-1}(((\1_n+\theta_2 A_a)^{-1}x)_i+(A_a^{-1}\mu)_i))} 
 \xi_a(x;\mu)\ ,\\
 (Z_\ib\xi_a)(x;\mu)& 
 = \xi_a(x+A_a^{-1}t_i;\mu-t_i)\ ,
 \end{split}
\end{equation*}
where the relation is 
\begin{equation*}
 Z_iZ_j=e^{-2\pi\sqrt{-1}\theta_a}Z_jZ_i\ ,\qquad 
 \theta_a
 =\bp \0_n & -(\1_n+\theta_2 A_a)^{-1}\theta_2 \\  
 \theta^t_2(\1_n+A_a\theta_2^t)^{-1} & \0_n \ep\ .
\end{equation*}
A constant curvature connection 
$\nabla_a: E_{g_a,\theta}\otimes\cL_\theta\to E_{g_a,\theta}$ 
is given as 
\begin{equation*}
 (\nabla_{a,1}\cdots \nabla_{a,2n})^t =  
 \bp
  \1_n &  \0_n \\ 
  \0_n & -(A_a^{-1}+\theta_2)^{-1} 
 \ep
 \bp \partial_x \\ 2\pi\sqrt{-1}x \ep\ ,
\end{equation*}
with its curvature 
\begin{equation*}
 F_a
 = \bp \0_n & (A_a^{-1}+\theta_2^t)^{-1} \\ 
   -(A_a^{-1}+\theta_2)^{-1} & \0_n \ep\ .
\end{equation*}
We assume that 
$A_{ab}$ is nondegenerate for any $a,b\in\Ob(\cCpre)$, $a\ne b$. 

For any $a,b\in\Ob(\cCpre)$, 
the space $\Hompre(a,b)$ is defined as follows. 
If $a\ne b$, 
$\Hompre(a,b):=\Hom(E_{g_a,\theta},E_{g_b,\theta})
=\cS(\R^n\times(\Z^n/A_{ab}\Z^n))$; 
the right action of $T_{\theta_a}$, 
generated by $U_i$ and $U_\ib$, 
and the left action of $T_{\theta_b}$, 
generated by $Z_i$ and $Z_\ib$, are defined by 
\begin{equation*}
  \begin{split}
 (\xi_{ab} U_i )(x;\mu)& = \xi_{ab}(x;\mu)
 e^{2\pi\sqrt{-1}((\1_n+\theta_2 A_a)^{-1}x)_i+(A_{ab}^{-1}\mu)_i))} \ ,\\
 (\xi_{ab} U_\ib)(x;\mu)& 
 = \xi_{ab}(x+(\1_n+\theta_2 A_b)A_{ab}^{-1}t_i;\mu-t_i)\ ,\\ 
 (Z_i \xi_{ab})(x;\mu)& 
 = e^{2\pi\sqrt{-1}(((\1_n+\theta_2 A_b)^{-1}x)_i+(A_{ab}^{-1}\mu)_i))} 
 \xi_{ab}(x;\mu)\ ,\\
 (Z_\ib \xi_{ab})(x;\mu)& 
 = \xi_{ab}(x+(\1_n+\theta_2 A_a)A_{ab}^{-1}t_i;\mu-t_i)\ .
 \end{split}
\end{equation*}
If $a=b$, then $\Hompre(a,b):=T_{\theta_a}=T_{\theta_b}$ and 
these actions are 
defined by the usual product of noncommutative torus 
$T_{\theta_a}=T_{\theta_b}$. 
Note that 
$\Hompre(a,b)=\Hom(E_{g_a,\theta},E_{g_b,\theta})$ 
is isomorphic to $E_{g_bg_a^{-1},\theta}$; 
for $a\ne b$, 
an element 
$\xi_{ab}\in\Hom(E_{g_a,\theta},E_{g_b,\theta})$ is identified with 
$\xi'_{ab}\in E_{g_bg_a^{-1},\theta}$ by 
the following relation: 
\begin{equation*}
 \xi'_{ab}(x',\mu)
 =\xi'_{ab}( (\1_n+\theta_2 A_a)^{-1}x,\mu) 
 =\xi_{ab}(x,\mu)\ .
\end{equation*}
A constant curvature connection 
$\nabla: \Hompre(a,b)\otimes\cL_\theta\to\Hompre(a,b)$ is given by 
\begin{equation*}
 (\nabla_{1}\cdots \nabla_{2n})^t=
 \bp
  \1_n &  \0_n \\ 
  \0_n & -(A_b^{-1}+\theta_2)^{-1}+(A_a^{-1}+\theta_2)
 \ep
 \bp \partial_x \\ 2\pi\sqrt{-1}x \ep\ 
\end{equation*}
if $a\ne b$, and if $a=b$, it is defined by the 
derivation $\nabla$ of noncommutative torus 
$T_{\theta_a}=T_{\theta_b}$ in eq.(\ref{delta-End}). 

For any $a,b,c\in\Ob(\cCpre)$ and 
$\xi_{ab}\in\Hompre(a,b)$, $\xi_{bc}\in\Hompre(b,c)$, the product 
$m:\Hompre(b,c)\otimes\Hompre(a,b)\to\Hompre(a,c)$ 
is given as follows:  

For $a=b$, it is the right action of $T_{\theta_a}=T_{\theta_b}$ 
on $\Hompre(b,c)$. 

For $b=c$, it is the left action of $T_{\theta_b}=T_{\theta_c}$ 
on $\Hompre(a,b)$. 

For $a=c$, the product 
$m:\Hompre(b,a)\otimes\Hompre(a,b)\to T_{\theta_a}$ is given by 
\begin{equation*}
 m(\xi_{ba},\xi_{ab})(x,\rho)
 =\sum_{\vec{m}\in\Z^n}\sum_{\mu\in\Z^n/A_{ab}\Z^n} 
 U_{\vec{m}}\int_{\R^n} dx^n 
 \xi_{ba}(x,\mu)\cdot (\xi_{ab}(x,-\mu)U_{-\vec{m}})\ .
\end{equation*}
For the remaining general case, it is given by 
\begin{equation*}
 \begin{split}
 & m(\xi_{bc},\xi_{ab})(x,\rho) = \\ 
 &\quad \sum_{u\in\Z^n}
  \xi_{bc}(x+(\1_n+\theta_2A_c)A_{bc}^{-1}(u-A_{ab}A_{ac}^{-1}\rho),
   -u+\rho)
  \cdot
  \xi_{ab}(x-(\1_n+\theta_2A_a)A_{ab}^{-1}(u-A_{ab}A_{ac}^{-1}\rho),u)\ .
 \end{split}
\end{equation*}
The trace map is then given by eq.(\ref{Tr-End}). 

\vspace*{0.2cm}

\noindent
{\bf Type $\theta_3$}\quad  
A category $\cCpre$ is constructed as follows. 
Any object $a\in\Ob(\cCpre)$ is associated with a pair 
$E_a:=(E_{g_a,\theta},\nabla_a)$. The basic module $E_{g_a,\theta}$ 
is defined by 
$$E_{g_a,\theta}=\cS(\R^n\times(\Z^n/A_a\Z^n))$$ 
for a nondegenerate symmetric matrix $A_a\in\Mat_n(\Z)$, where 
\begin{equation*}
 g_a= \bp \1_{2n} & \0_{2n} \\ F_{a,0} & \1_{2n}\ep\ ,\qquad 
 F_{a,0} =\bp \0_n & A_{a}\\ -A_{a} & \0_n\ep\ .
\end{equation*}
For $\xi_a\in E_{g_a,\theta}$, 
the action of each generator is defined by 
\begin{equation*}
 \begin{split}
 (\xi_aU_i)(x;\mu)& 
 =\xi_a(x;\mu) e^{2\pi\sqrt{-1}(x_i+(A_a^{-1}\mu)_i))} \ ,\\
 (\xi_aU_\ib)(x;\mu)& 
 =\xi_a(x+A_a^{-1}t_i;\mu-t_i) e^{-\pi\sqrt{-1}x^tA_a\theta_3 t_i}
 \ ,\qquad i=1,\cdots, n \ , 
 \end{split}
\end{equation*} 
and the endomorphisms are generated by 
\begin{equation*}
 \begin{split}
 (Z_i\xi_a)(x;\mu)& 
 = e^{2\pi\sqrt{-1}(x_i+(A_a^{-1}\mu)_i))} \xi_a(x;\mu)  \ ,\\
 (Z_\ib \xi_a)(x;\mu)& 
 = e^{\pi\sqrt{-1}x^tA_a\theta_3 t_i}\xi_a(x+A_a^{-1}t_i;\mu-t_i) \ .
 \end{split}
\end{equation*} 
A constant curvature connection 
$\nabla_a: E_{g_a,\theta}\otimes\cL_\theta\to E_{g_a,\theta}$ 
is given as 
\begin{equation*}
 (\nabla_{a,1}\cdots \nabla_{a,2n})^t = 
 \bp
  \1_n &  -\ov{2}A_a\theta_3 A_a \\ 
  \0_n & -A_a
 \ep
 \bp \partial_x \\ 2\pi\sqrt{-1}x \ep\ ,
\end{equation*}
with its curvature 
\begin{equation*}
 F_a = 
  \bp -A_a\theta_3 A_a & A_a \\ 
       -A_a         & \0_n \ep\ .
\end{equation*}
We assume that 
$A_{ab}$ is nondegenerate for any $a,b\in\Ob(\cCpre)$, $a\ne b$. 

For any $a,b\in\Ob(\cCpre)$, 
the space $\Hompre(a,b)$ is defined as follows. 
If $a\ne b$, 
$\Hompre(a,b):=\Hom(E_{g_a,\theta},E_{g_b,\theta})
=\cS(\R^n\times(\Z^n/A_{ab}\Z^n))$; 
the right action of $T_{\theta_a}$, 
generated by $U_i$ and $U_\ib$, 
and the left action of $T_{\theta_b}$, 
generated by $Z_i$ and $Z_\ib$, are defined by 
\begin{equation*}
  \begin{split}
 (\xi_{ab} U_i)(x;\mu)& 
 =\xi_{ab}(x;\mu) e^{2\pi\sqrt{-1}(x_i+(A_{ab}^{-1}\mu)_i))} \ ,\\
 (\xi_{ab} U_\ib)(x;\mu)& 
 =\xi_{ab}(x+A_{ab}^{-1}t_i;\mu-t_i) 
 e^{-\pi\sqrt{-1}x^tA_{ab}\theta_3 t_i}\ ,\\
 (Z_i \xi_{ab})(x;\mu)& 
 = e^{2\pi\sqrt{-1}(x_i+(A_{ab}^{-1}\mu)_i))} \xi_{ab}(x;\mu)  \ ,\\
 (Z_\ib \xi_{ab})(x;\mu)& 
 = e^{\pi\sqrt{-1}x^tA_{ab}\theta_3 t_i} 
 \xi_{ab}(x+A_{ab}^{-1}t_i;\mu-t_i) \ .
 \end{split}
\end{equation*}
If $a=b$, then $\Hompre(a,b):=T_{\theta_a}=T_{\theta_b}$ and 
the left and right actions on it are 
defined by the usual product of noncommutative torus 
$T_{\theta_a}=T_{\theta_b}$. 

A constant curvature connection 
$\nabla_a: \Hompre(a,b)\otimes\cL_\theta\to\Hompre(a,b)$ is given by 
\begin{equation*}
 (\nabla_{1}\cdots \nabla_{2n})^t = 
 \bp \1_n & -\ov{2}A_{ab}^+\theta_3 A_{ab} \\ 
     \0_n & -A_{ab} \ep
 \bp \partial_x \\ 2\pi\sqrt{-1}x \ep\ 
\end{equation*}
if $a\ne b$, and if $a=b$, it is defined by the usual
derivation $\nabla$ of noncommutative torus 
$T_{\theta_a}=T_{\theta_b}$ (\ref{delta-End}). 

For any $a,b,c\in\Ob(\cCpre)$ and 
$\xi_{ab}\in\Hompre(a,b)$, $\xi_{bc}\in\Hompre(b,c)$, 
the product $m:\Hompre(b,c)\otimes\Hompre(a,b)\to\Hompre(a,c)$ 
is given as follows:  

For $a=b$, it is the right action of $T_{\theta_a}=T_{\theta_b}$ 
on $\Hompre(b,c)$. 

For $b=c$, it is the left action of $T_{\theta_b}=T_{\theta_c}$ 
on $\Hompre(a,b)$. 

For $a=c$, the product 
$m:\Hompre(b,a)\otimes\Hompre(a,b)\to T_{\theta_a}$ is given by 
\begin{equation*}
 m(\xi_{ba},\xi_{ab})(x,\rho)
 =\sum_{\vec{m}\in\Z^n}\sum_{\mu\in\Z^n/A_{ab}\Z^n} 
 U_{\vec{m}}\int_{\R^n} dx^n 
 \xi_{ba}(x,\mu)\cdot (\xi_{ab}(x,-\mu)U_{-\vec{m}})\ .
\end{equation*}
For the remaining general case, it is given by 
\begin{equation*}
 m(\xi_{bc},\xi_{ab})(x,\rho)
 =\sum_{u\in\Z^n}
  \xi_{bc}(x',-u+\rho)\cdot
  \exp\(-\pi\sqrt{-1}{x'}^tA_{bc}\theta_3 A_{ab}x''\)\cdot
  \xi_{ab}(x'',u)\ ,
\end{equation*}
where $x':=x+A_{bc}^{-1}(u-A_{ab}A_{ac}^{-1}\rho)$ and 
$x'':=x-A_{ab}^{-1}(u-A_{ab}A_{ac}^{-1}\rho)$. 

The trace map is then given by eq.(\ref{Tr-End}).

By direct calculations, one obtains the followings. 
\begin{lem}
For a fixed noncommutative parameter of type $\theta_1$, $\theta_2$ or 
$\theta_3$, 

 (Associativity):\ the composition $m$ of morphisms is associative. 

 (Leibniz rule):\ the constant curvature connections on morphisms 
satisfy the Leibniz rule. 
\qed\end{lem}
Then, in any case of the Type $\theta_s$, $s=1,2,3$, 
$\cCpre$ forms a category in Definition \ref{defn:pre-C} and 
then the corresponding category $\cC$ forms a cyclic CDG category. 

In particular, by looking at the condition that $d:V_{ab}\to V_{ab}$ 
satisfies $d^2=0$ explicitly, one can see the followings: 
\begin{prop}
For Type $\theta_1$, Type $\theta_2$ such that $\theta_2^t=-\theta_2$ 
and Type $\theta_3$, 
two objects $a,b\in\Ob(\cC)$ in the CDG category $\cC$ 
forms a full sub-DG category $\cC^\fh$ of $\cC$ 
for some $\fh\in\Lambda^2\subset V^2$ 
if and only if 
\begin{equation}
 A_a\theta_s A_a=A_b\theta_s A_b\ ,\qquad s=1,2,3\ 
 \label{cd}
\end{equation}
holds. 
 \label{prop:cd}
\qed\end{prop}
Thus, we have obtained the cyclic DG categories $\cC^\fh$
on noncommutative tori with three types 
of noncommutativities $\theta_1$, $\theta_2$ and $\theta_3$.

Let us calculate the compositions of the 
morphisms of the zero-th cohomologies 
$H^0(\cV):=\oplus_{a,b\in\Ob(\cC^\fh)} H^0(V_{ab})$ 
of these DG categories $\cC^\fh$. 
They define ring structures on $H^0(\cV)$, which are sub-rings of 
the full cohomologies $H^*(\cV)$. 
We shall observe that the ring $H^0(\cV)$ actually depends on 
the noncommutative parameter $\theta_s$, 
as opposed to the complex one-tori case 
\cite{foliation, KimKim, PoSc, nchms}. 
We remark that, 
from a homotopy algebraic viewpoint, 
a DG category is homotopy equivalent to 
a {\em minimal} $A_\infty$ category. 
Then, by forgetting the higher compositions of the 
minimal $A_\infty$ category, 
one obtains the ring $H^*(\cV)$. 
Thus, if at least the sub-ring $H^0(\cV)$ is deformed, 
the minimal $A_\infty$-structure is also deformed.

\vspace*{0.3cm}

Recall that $H^0(V_{ab})$ is given by 
$\Ker(d:V_{ab}^0\to V_{ab}^1)=
\cap_{i=1}^n\Ker(\nabb_i:V_{ab}^0\to V_{ab})$. 

\vspace*{0.2cm}

\noindent
{\bf Type $\theta_1$}\quad
For $a,b\in\Ob(\cCpre)$, $a\ne b$, 
the holomorphic structure $\nabb_i:\Hompre(a,b)\to\Hompre(a,b)$ 
is given by 
\begin{equation*}
 (\nabb_1\cdots\nabb_n)^t =
 \left(\1_n + \frac{\sqrt{-1}}{2}A_{ab}^+\theta\right)\partial_x 
     +2\pi A_{ab}x\ .
\end{equation*}
The cohomology $H^0(V_{ab})$ is spanned by the bases 
of the form 
\begin{equation}\label{theta-vector_1}
 e_{ab}^\mu(x;\rho)
 :=C_{ab}\delm{A_{ab}}^\mu_\rho\exp\( -\pi\( x^t M_{ab} x \) \)\ ,
 \qquad \mu\in\Z^n/A_{ab}\Z^n\ ,
\end{equation}
where 
$C_{ab}\in\C$ is an appropriate rescaling and 
$M_{ab}\in\Mat_n(\C)$ should be a symmetric matrix 
satisfying 
\begin{equation}
 (\nabb_1\cdots\nabb_n)\, 
 \left(\exp\( -\pi\( x^t M_{ab} x \) \)\right)=0\ .
 \label{theta_1-cd}
\end{equation}
The condition (\ref{theta_1-cd}) turns out to be 
\begin{equation*}
 -\left(\1_n+\frac{\sqrt{-1}}{2}A_{ab}^+\theta_1\right)M_{ab} +A_{ab}=0\ .
\end{equation*}
This $M_{ab}$ is symmetric if and only if the condition (\ref{cd})
holds: 
\begin{equation*}
 A_a\theta_1 A_a=A_b\theta_1 A_b\ ,
\end{equation*}
and then the explicit form of $M_{ab}$ is given by 
\begin{equation*}
 M_{ab}=A_{ab}
 \(A_{ab}+\frac{\sqrt{-1}}{2}(A_a\theta_1 A_b-A_b\theta_1 A_a)\)^{-1}A_{ab}\ .
\end{equation*}
Here the real part of $M_{ab}\in\Mat_n(\C)$ should be positive
definite in order for $e_{ab}^\mu$ to exist in $H^0(V_{ab})$. 
Note that the part of $M_{ab}$ is positive definite if and only if 
$A_{ab}$ is positive definite. 
This $e_{ab}^\mu\in H^0(V_{ab})$ in eq.(\ref{theta-vector_1}) 
is a {\em theta vector}. 
Thus, for $a,b\in\Ob(\cC)$ such that $A_a\theta_1 A_a=A_b\theta_1 A_b$ 
and $A_{ab}$ is positive definite, 
$\dim(H^0(V_{ab}))=\sharp(\Z^n/A_{ab}\Z^n)=\det(A_{ab})$. 
For the rescaling $C_{ab}$ in eq.(\ref{theta-vector_1}), 
we set 
\begin{equation*}
 C_{ab}:=
 \frac{\det(\1_n+\i A_a\theta)^{\ov{4}}
\det(\1_n+\i A_b\theta)^{\ov{4}}}
{\det(\1_n+\2i A_{ab}^+\theta)^{\ov{2}}}\ .
\end{equation*}
Assume now that $A_{ab}$ and $A_{bc}$ are positive definite. 
Then we get the product formula: 
\begin{equation*}
 \begin{split}
 & m(e^\mu_{ab},e^\nu_{bc})
 =\frac{}{}
 \sum_{\rho\in\Z^n/A_{ac}\Z^n}
 \sum_{u\in\Z^n}\delm{A_{ab}}^\mu_{-u+\rho}\delm{A_{bc}}^\nu_u \\
 &\quad \exp\(-\pi(u-A_{bc}A_{ac}^{-1}\rho)^t
 \((A_{ab}^{-1}+A_{bc}^{-1})(\1_n+\sqrt{-1}A_b\theta)^{-1}\)
 (u-A_{bc}A_{ac}^{-1}\rho)\)
 \cdot e_{ac}^\rho\ .
 \end{split}
\end{equation*}
Note that the $n$ by $n$ matrix 
$(A_{ab}^{-1}+A_{bc}^{-1})(\1_n+\sqrt{-1}A_b\theta)^{-1}$ 
is automatically symmetric due to the condition 
$A_a\theta_1 A_a=A_b\theta_1 A_b=A_c\theta_1 A_c$.

In this Type $\theta_1$ case, these theta vectors $\{e_{ab}^\mu\}$ 
can be described by theta functions, 
where the product of two theta vectors just corresponds to 
the Moyal product of two theta functions \cite{nctheta}.

\vspace*{0.2cm}

\noindent
{\bf Type $\theta_2$}\quad
For $a,b\in\Ob(\cCpre)$, $a\ne b$, 
the holomorphic structure $\nabb_i:\Hompre(a,b)\to\Hompre(a,b)$ 
is given by 
\begin{equation*}
  (\nabb_1\cdots\nabb_n)^t = 
 \partial_x +2\pi
 \((A_b^{-1}+\theta_2)^{-1}-(A_a^{-1}+\theta_2)^{-1}\)x\ . 
\end{equation*}
The theta vectors are of the form 
\begin{equation*}
 e_{ab}^\mu(x,\rho)=\delm{A_{ab}}^\mu_\rho\, \exp\(-\pi x^t M_{ab}
x\)\ ,\qquad \mu\in\Z^n/A_{ab}\Z^n\ , 
\end{equation*}
where $M_{ab}\in\Mat_n(\R)\subset\Mat_n(\C)$ is given by 
\begin{equation}\label{Mab-theta_2}
 M_{ab}:=(A_b^{-1}+\theta_2)^{-1}-(A_a^{-1}+\theta_2)^{-1}
=(\1_n+A_b\theta_2)^{-1}A_{ab}(\1_n+\theta_2A_a)^{-1} \ ,
\end{equation}
which should be a symmetric positive definite matrix. 
Then, one has $\dim(H^0(V_{ab}))=\det(A_{ab})$.

Assume that 
$M_{ab}$ and $M_{bc}$ are positive definite. 
Then, $M_{ac}$ is also positive definite. 
The product of $e_{ab}^\mu$ with $e_{bc}^\nu$ is then 
\begin{equation*}
 \begin{split}
 & m(e_{ab}^\mu,e_{bc}^\nu)
 = \sum_{\rho\in\Z^n/A_{ac}\Z^n}
 \sum_{u\in\Z^n}\delm{A_{ab}}^\mu_{-u+\rho}\delm{A_{bc}}^\nu_u \\
 & \exp\(-\pi(u-A_{bc}A_{ac}\rho)^t
((A_{ab}^{-1}+A_{bc}^{-1})(\1_n+A_b\theta_2^t)(\1_n+A_b\theta_2)^{-1})
 (u-A_{bc}A_{ac}\rho)\)\cdot e_{ac}^\rho\ .
 \end{split}
\end{equation*}
In particular, one can see that the structure constant does not depend
on $\theta_2$ if and only if $theta_2$ is symmetric: 
$\theta_2^t=\theta_2$. 
This gives the reason that the structure constant of the 
product does not depend on the noncommutative parameter in the 
case of noncommutative real two-tori \cite{foliation,PoSc,nchms}. 
See also \cite{KimLee,KimKim2}, where 
for a complex two-tori with 
noncommutativity of Type $\theta_2$ with symmetric $\theta_2$, 
such structure constants are computed and checked to be 
independent of the noncommutativity $\theta_2$.

When $\theta_2$ is antisymmetric, $M_{ab}$ in eq.(\ref{Mab-theta_2}) 
is symmetric if and only if $A_a\theta_2 A_a= A_b\theta_2 A_b$, 
where the $n$ by $n$ matrix 
$(A_{ab}^{-1}+A_{bc}^{-1})(\1_n+A_b\theta_2^t)(\1_n+A_b\theta_2)^{-1}
\in\Mat_n(\C)$ is symmetric: 
\begin{equation*}
 (A_{ab}^{-1}+A_{bc}^{-1})(\1_n+A_b\theta_2^t)(\1_n+A_b\theta_2)^{-1}
 =(\1_n-\theta_2 A_b)^{-1}(\1_n-\theta_2^tA_b)(A_{ab}^{-1}+A_{bc}^{-1})\ .
\end{equation*}

\vspace*{0.2cm}

\noindent
{\bf Type $\theta_3$}\quad
For $a,b\in\Ob(\cCpre)$, $a\ne b$, 
the holomorphic structure $\nabb_i:\Hompre(a,b)\to\Hompre(a,b)$ 
is given by 
\begin{equation*}
 (\nabb_1\cdots\nabb_n)^t = 
 \partial_x + 2\pi\(\1_n-\frac{\sqrt{-1}}{2}A_{ab}^+\theta_3\)A_{ab} x
 \ . 
\end{equation*}
The theta vectors are of the form 
\begin{equation*}
 e_{ab}^\mu(x,\rho)=\delm{A_{ab}}^\mu_\rho\, \exp\(-\pi x^t M_{ab}
x\)\ ,\qquad \mu\in\Z^n/A_{ab}\Z^n\ ,
\end{equation*}
where $M_{ab}\in\Mat_n(\C)$ is given by 
\begin{equation*}
 M_{ab}:=\(\1_n-\frac{\sqrt{-1}}{2}A_{ab}^+\theta_3\)A_{ab}\ ,
\end{equation*}
which should be a symmetric matrix whose real part is positive definite. 
Here, again, the real part of $M_{ab}$ is positive definite 
if and only if $A_{ab}$ is positive definite. 
Then, one has $\dim(H^0(V_{ab}))=\det(A_{ab})$. 
The condition that $M_{ab}$ above is symmetric is equal to 
\begin{equation*}
 A_{ab}^+\theta_3A_{ab}=(A_{ab}^+\theta_3A_{ab})^t\ ,
\end{equation*}
which is in fact equivalent to $A_a\theta_3 A_a=A_b\theta_3 A_b$.

Assume now that $A_{ab}$ and $A_{bc}$ are positive definite. 
The product of two theta vectors is given by 
\begin{equation*}
 \begin{split}
 & m(e_{ab}^\mu, e_{bc}^\nu)
  =\sum_{\rho\in\Z^n/A_{ac}\Z^n}
 \sum_{u\in\Z^n}\delm{A_{ab}}^\mu_{-u+\rho}\delm{A_{bc}}^\nu_{u} \\ 
 &\quad 
 \exp\(-\pi(u-A_{bc}A_{ac}^{-1}\rho)^t
 \((A_{ab}^{-1}+A_{bc}^{-1})(\1_n-\sqrt{-1}A_b\theta_3)\)
 (u-A_{bc}A_{ac}^{-1}\rho)\)
 \cdot e_{ac}^\rho\ .
 \end{split}
\end{equation*}
One can show that the matrix 
$(A_{ab}^{-1}+A_{bc}^{-1})(\1_n-\sqrt{-1}A_b\theta_3)$ defining a
quadratic form in the expression above is symmetric: 
\begin{equation*}
 (A_{ab}^{-1}+A_{bc}^{-1})(\1_n-\sqrt{-1}A_b\theta_3)
 =(\1_n+\sqrt{-1}\theta_3 A_b)(A_{ab}^{-1}+A_{bc}^{-1})\ .
\end{equation*}
Thus, we have seen that the structure constants 
$C_{abc,\rho}^{\mu\nu}$ depend on the noncommutative parameter 
$\theta$ in all these three cases.

Though in this paper we have fixed a constant curvature connection 
for a Heisenberg module, 
we can also take all the constant curvature connections on a 
Heisenberg module into account in a similar way as in the
noncommutative complex one-tori case \cite{foliation,nchms}. 
It might also be interesting to investigate the details of the moduli 
space of the constant curvature connections on a Heisenberg module 
on $T^{2n}_\theta$ which is known to form a commutative torus $T^{2n}$.

We end with showing an example for the case of noncommutative 
complex two-torus ($n=2$). In this case, 
for any fixed 
$\theta_s$, 
$s=1,2,3$, 
the condition eq.(\ref{cd}) reduces to 
\begin{equation*}
 \det(A_a)=\det(A_b)\ .
\end{equation*}
Thus, for the objects of $\cC^f$, one can in general 
have infinite number of objects. 
For instance, diagonal matrices $A\in\Mat_n(\Z)$ with 
$\det(A)=-4$ are 
\begin{equation*}
 A_a=\bp 1 & 0 \\ 0 & -4\ep\ ,\quad 
 A_b=\bp 2 & 0 \\ 0 & -2\ep\ ,\quad 
 A_c=\bp 4 & 0 \\ 0 & -1\ep\ ,
\end{equation*}
and $A_{a'}:=-A_a$, $A_{b'}:=-A_b$, $A_{c'}:=-A_c$. 
Since the zero-th cohomologies of morphisms between 
$\{a,b,c\}$ and $\{a',b',c'\}$ are absent, 
let us concentrate on the one side $\{a,b,c\}$. 
Then, all $A_{ab}, A_{bc}$ and $A_{ac}$ are positive definite, 
and the dimensions of the zero-th cohomologies are 
\begin{equation*}
 \dim(H^0(V_{ab}))=2\ ,\qquad 
 \dim(H^0(V_{bc}))=2\ ,\qquad 
 \dim(H^0(V_{ac}))=9\ .
\end{equation*}
However, there exist infinite symmetric matrices $A\in\Mat_n(\Z)$ 
with $\det(A)=-4$, 
since 
the matrix $g^tA g$ has $\det(A)=-4$ for any $SL(2,\Z)$ element $g$.

\end{document}